\documentclass[journal]{IEEEtran}
\usepackage{braket}
\usepackage{amsmath}
\usepackage{amssymb}
\usepackage{graphicx}
\usepackage{cite}
\usepackage{epstopdf}
\usepackage{mathtools}
\usepackage{url}
\usepackage[colorlinks,linkcolor=blue,anchorcolor=blue,citecolor=blue]{hyperref}
\usepackage{winsnotation}
\usepackage{bbm}
\usepackage{caption}
\usepackage{tikz,xcolor,hyperref}

\begin{document}

\newcommand{\tr}{\text{Tr}}
\newcommand{\ketbra}[1]{\ket{#1}\bra{#1}}
\newcommand{\bigchi}{\makebox{\large\ensuremath{\chi}}}
\newcommand\blfootnote[1]{%
  \begingroup
  \renewcommand\thefootnote{}\footnote{#1}%
  \addtocounter{footnote}{-1}%
  \endgroup
}

\newcommand{\overbar}[1]{\mkern 3.7mu\overline{\mkern-3.7mu#1\mkern-3.7mu}\mkern 3.7mu}

%\history{Date of publication xxxx 00, 0000, date of current version xxxx 00, 0000.}
%\doi{10.1109/TQE.2020.DOI}

\title{Enhanced Uplink Quantum Communication with Satellites via Downlink Channels}

\author{
\IEEEauthorblockN{Eduardo Villase\~nor$^1$, Mingjian He$^1$, Ziqing Wang$^1$, Robert Malaney$^1$ and Moe Z. Win$^2$.}\\
	\IEEEauthorblockA{${}^1$School of Electrical Engineering  \& Telecommunications,\\
		The University of New South Wales,  Sydney, NSW 2052, Australia.\\
		${}^2$Laboratory for Information and Decision Systems, Massachusetts Institute of Technology, \\ Cambridge, MA 02139, USA.}
}
\maketitle

%\author{
%	\uppercase{Eduardo Villase\~nor}\insertorcid{\orcidauthorE}\authorrefmark{1}, \IEEEmembership{Student Member, IEEE},
%	\uppercase{Mingjian He}\insertorcid{\orcidauthorM}\authorrefmark{1}, \IEEEmembership{Student Member, IEEE},
%	\uppercase{Ziqing Wang}\insertorcid{\orcidauthorZ}\authorrefmark{1}, \IEEEmembership{Student Member, IEEE},
%	\uppercase{Robert Malaney}\insertorcid{\orcidauthorR}\authorrefmark{1}, \IEEEmembership{Senior Member, IEEE},
%	and \uppercase{Moe Z. Win}\insertorcid{\orcidauthorW}\authorrefmark{2}, \IEEEmembership{Fellow, IEEE}}
%
%
%\address[1]{School of Electrical Engineering and Telecommunications, the University of New South Wales, Sydney, NSW 2052, Australia.}
%\address[2]{Laboratory for Information and Decision Systems, Massachusetts Institute of Technology, Cambridge, MA 02139, USA.}

%\tfootnote{This paragraph of the first footnote will contain support
%information, including sponsor and financial support acknowledgment}

%\markboth
%{Villase\~nor \headeretal: Enhanced Uplink Quantum Communication with Satellites via Downlink Channels}
%{Villase\~nor \headeretal: Enhanced Uplink Quantum Communication with Satellites via Downlink Channels}

%\corresp{Corresponding author: R. Malaney (email: r.malaney@unsw.edu.au).}

\begin{abstract}
In developing the global Quantum Internet, quantum
communication with low-Earth-orbit  satellites will
play a pivotal role. Such communication will need to be two way:
effective not only in the satellite-to-ground (downlink)
channel but also in the ground-to-satellite channel (uplink). Given
that losses on this latter channel are significantly larger
relative to the former, techniques that can exploit the superior
downlink to enhance quantum communication in
the uplink should be explored. In this work we do just that -
exploring how continuous variable entanglement in the form
of two-mode squeezed vacuum (TMSV) states can be used to
significantly enhance the fidelity of ground-to-satellite quantum-state
transfer relative to direct uplink-transfer. More specifically,
through detailed phase-screen simulations of beam evolution
through turbulent atmospheres in both the downlink
and uplink channels, we demonstrate how a TMSV teleportation
channel created by the satellite can be used to dramatically
improve the fidelity of uplink coherent-state transfer relative
to direct transfer. We then show how this, in turn, leads to
the uplink-transmission of a higher alphabet of coherent states.
Additionally, we show how non-Gaussian operations acting on
the received component of the TMSV state at the ground station
can lead to even further enhancement. Since TMSV states can
be readily produced in situ on a satellite platform and
form a reliable teleportation channel for most
quantum states, our work suggests future satellites forming
part of the emerging Quantum Internet should be designed with
uplink-communication via TMSV teleportation in mind.
\end{abstract}

%\begin{keywords}
%Quantum communications, quantum teleportation, satellites, free-space-optics.
%\end{keywords}

%\vspace*{-1.5cm}

\bstctlcite{IEEEexample:BSTcontrol} % set references style
\bibliographystyle{IEEEtran}

\section{Introduction}
\IEEEPARstart{Q}{uantum communications}
via low-Earth-orbit (LEO) represent a critical component  of the so-called Quantum Internet - a new heterogeneous global communication system based on classical and quantum communication techniques
whose information security will be underpinned by quantum protocols such as quantum key distribution (QKD)\blfootnote{Corresponding author: R. Malaney (email: r.malaney@unsw.edu.au)}.

This new internet will also be used as the backbone communication system inter-connecting future quantum computers via routed quantum information transfer.
The Quantum Internet paradigm has taken large steps forward in the past few years, particularly with the spectacular success of Micius - the first quantum-enabled satellite launched in 2016 \cite{yin2017, liao2017satellite, Ren2017, 10.1103/PhysRevLett.120.030501}. Building on the pioneering Micius mission, some twenty-plus satellite missions are now under development \cite{Bedington2017} - some at the advanced design phase.

 The importance of satellite-based  technology to the Quantum Internet paradigm lies in a satellite's ability
 %However, achieving practical and global quantum communications involves monumental technological challenges \cite{QKDReview}.
 to transmit quantum signals through much longer distances relative to terrestrial-only links \cite{yin2017, liao2017satellite, QKDReview}.
 %The use of satellites in space to receive and transmit quantum signals has arisen as a prominent solution to achieve global quantum communications.
Indeed, the Micius experiment has demonstrated quantum communication over a range of 7,600km \cite{10.1103/PhysRevLett.120.030501} - a feat put into perspective by the current terrestrial-only quantum communication record of 500km \cite{PhysRevLett.124.070501}.

The Micius experiment deployed
quantum communication protocols via discrete variable (DV)  technology where the quantum information was encoded in the  polarization state of single photons  \cite{liao2017satellite, Ren2017}.
Alternatively, continuous variables (CV) quantum information, where the information is encoded in the quadratures of the electromagnetic field of optical states, is widely touted as perhaps a more promising candidate to transfer quantum information \cite{.5kmCVQKD, Valivarthi2020}. This is largely
due to the relative technical simplicity (and maturity) of the CV-enabled devices required to send, receive, and measure quantum signals, robustness against background noise, and the potential of the enlarged Hilbert space associated with CV systems to lead to enhanced communication throughput in practical settings.\footnote{In theory both DV and CV communication deliver the same throughput, and the reality is both systems have their pros and cons. However, there certainly is a school-of-thought that in many pragmatic systems, the higher-dimensional encoding space directly available to CV systems will lead to enhanced outcomes. A detailed discussion of the pros and cons of both  DV and CV systems is given in \cite{CVDV}.}
%This is specially the case in the context of CV-QKD protocols \cite{.5kmCVQKD}.
For these reasons, there is great interest in pursuing designs of CV-enabled quantum satellites, with many recent studies focusing on the more feasible  satellite-to-ground (downlink) transmission of quantum signals, largely with a view to enable CV-QKD \cite{NedaReview, FeasibilityDownlinkCVQKD}. As yet, there has not been any experimental realizations of satellite-based CV quantum communications. In this work, we turn to a hitherto overlooked type of satellite-based CV quantum communications, namely, the use of CV quantum downlink communications as a means to enhance ground-to-satellite (uplink) quantum communications with a LEO satellite.

The main challenge faced in satellite-based quantum communications is the degradation of the signal as it is transmitted through the turbulent atmosphere of Earth \cite{AtmosphericChannels, AtmosphericQKD, PhysRevA.99.053830}, a degradation that is almost always larger than  the noise introduced by the components used \cite{PracticalCVQKD}.
It is well documented that uplink satellite laser communications is considerably more challenging compared to downlink satellite transmission: the turbulent eddies in the Earth's atmosphere have a more disruptive effect in the uplink channel. This is because the size of the eddies
encountered by a laser beam in the downlink at the atmospheric entry point
are significantly larger than the laser beam's transverse dimensions (spot size) at the entry point, whereas in  the uplink the opposite is true \cite{andrews_book1}. The consequence of this is that an asymmetry in the channels exists, with the uplink beam profile evolving in a more random fashion, especially in regard to beam wandering effects. Ultimately, this asymmetry manifests itself  in higher losses in the uplink channel \cite{andrews_book1,AOUplink}.

Here, we investigate the use of quantum resources delivered through the satellite downlink channel as a resource for teleportation in the uplink, and the subsequent use of that teleportation resource to enhance quantum communications relative to simple direct uplink transmission.
%CV quantum teleportation has been demonstrated using optical fibre over 6 km \cite{TeleportationExperiment}, although a demonstration of teleportation via the free-space optical channel is still pending.
%In quantum teleportation a bipartite $AB$ entangled resource state is required, where part $A$ is combined to the input state, while part $B$ is sent trough the noise channel \cite{braustein1998teleportationCV}. In this manner, by means of a projective measurement on part $A$ and the input state, and a corrective operation on $B$ the input state is recovered as the final output.
More specifically, we consider the use of a two-mode-squeezed vacuum (TMSV)  quantum teleportation channel created via the downlink channel as a resource to teleport a coherent state from the  ground station to a LEO satellite.
%we will call this channel the {\it upload} channel.
%We compare the resulting fidelity achieved via this \emph{indirect} teleportation channel with the fidelity obtained via direct transmission of the coherent state in the uplink channel.
We will see  that for uplink communications, the use of the teleportation channel leads to significantly higher fidelities compared to the direct transmission. Moreover, we find that the teleportation channel is capable of transferring coherent states with larger amplitudes, something that is very difficult via direct transmission. This latter attribute is important for many CV-based quantum protocols, such as CV-QKD, since for these protocols the capability to transmit coherent states of different amplitudes is a key requirement.

The main contributions of this work can be summarized thus. (i) Through a series of detailed phase-screen simulations we quantify the asymmetric losses experienced by the downlink and uplink channels of a LEO satellite in quantum communication with a terrestrial ground station. Moreover, we expand previous analyses of the uplink and downlink channels by quantifying and including the excess noise that arises from each channel.
This excess noise limits the accuracy of the quadrature measurements, effectively reducing the amount of transferred quantum information. (ii) Using these same simulations we then determine the fidelity of coherent state transfer through direct uplink transfer. (iii) We model the creation of a resource CV teleportation channel in the downlink created by sending from the satellite one mode of an in situ produced TMSV state. (iv) We then use that resource to determine the fidelity of coherent state transfer  to the satellite via teleportation, quantifying the gain achieved over direct transfer. (v) We then investigate a series of non-Gaussian operations that can be invoked on the received TMSV mode at the ground station as a means to further enhance uplink coherent-state transfer via teleportation. Specifically, we investigate photon subtraction, addition, and catalysis  as the non-Gaussian operations - identifying the gains in teleportation fidelity achieved for each scheme. Sequences of these non-Gaussian operations are also investigated, and the optimal scheme amongst them identified.

The remainder of this paper is as follows. In section~II we describe CV teleportation through noisy channels. In section~III we  detail our phase screen simulations, comparing their predictions with a range of theoretical models, and discussing the implications of our simulations in the context  asymmetric downlink/uplink channel losses. In section~IV we discuss a series of non-Gaussian operations that can be applied to a TMSV state, discussing  their roles in potentially enhancing CV teleportation via a noisy TMSV channel. In section V we discuss application of our schemes to a wider range of states,
and discus differences with the DV-only scheme of Micius. We draw our conclusions in section VI.

{\it Notation:}
% Random variables are displayed in sans serif, upright fonts; their realizations in serif, italic fonts.
Operators are denoted by uppercase letters.
% For example, a random variable and its realization are denoted by x and x ; a random operator and its realization are denoted by X and X , respectively.
The sets of complex numbers and of positive integer numbers are denoted by $\mathbb{C}$ and $\mathbb{N}$, respectively. For $z \in \mathbb{C}$: $|z |$ and $\arg(z)$ denote the absolute value and the phase, respectively; $\text{Re}(z)$ and $\text{Im}(z)$ denote the real part and the imaginary part, respectively; $z^*$ is the complex conjugate;
and $i = \sqrt{-1}$. The trace and the adjoint of an operator are denoted by $\tr\{\cdot\}$ and $(\cdot)^\dagger$, respectively. The annihilation, the creation, and the identity operators are denoted by $\M{A}$, $\M{A}^\dagger$,
and $\M{I}$, respectively. The displacement operator with parameter $\alpha \in \mathbb{C}$ is $\M{D}(\alpha) = \exp[\alpha \M{A}^\dagger - \alpha^* \M{A} ].$
% The rotation operator with parameter �� �� R is R�� = exp{�|?��A��A}.
% The Kronecker delta function, the discrete delta function, and the generalized delta function are denoted by ��n,m , ��n, and ��(?), respectively.

%%%%%%%%%%%%%%%%%%%%%%%%%%%%%%%%%%%%%%%%%%%%%%%%%%%%%%%%%%%
%%%%%%%%%%%%%%%%%%%%%%%%%%%%%%%%%%%%%%%%%%%%%%%%%%%%%%%%%%%
%%%%%%%%%%%%%%%%%%%%%%%%%%%%%%%%%%%%%%%%%%%%%%%%%%%%%%%%%%%
\section{Continuous variable teleportation}
\begin{figure}
\centering
\includegraphics[width=.48\textwidth]{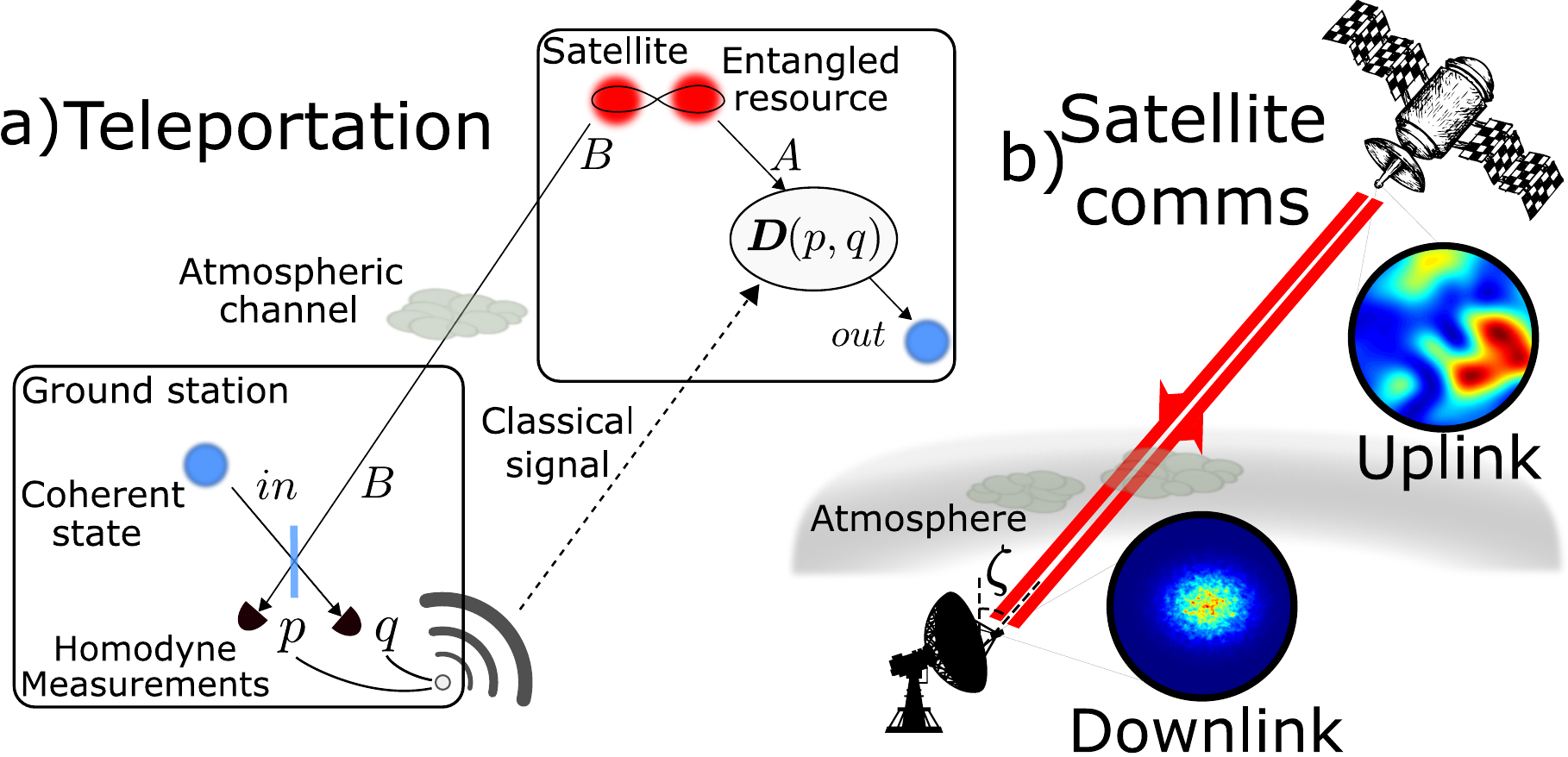}
\caption{a) CV  teleportation of a coherent state using a bipartite entangled resource between a satellite and a ground station. Homodyne measurement results are transmitted by the ground station after combining the received quantum signal with the coherent state. The satellite uses the measurement results to apply a displacement operator on the remaining mode of the entangled state to obtain the teleported state. b) In satellite communications the downlink channel is considerably less noisy than the uplink channel.}
\label{fig:drawing}
\end{figure}

We consider the teleportation protocol introduced in \cite{braustein1998teleportationCV}. Here, we are considering the parties involved in the teleportation are a ground station and a satellite in space, with the quantum channel between them corresponding to the free-space atmospheric channel as exemplified in Fig. \ref{fig:drawing}a). The teleportation protocol starts with the generation of a bipartite entangled resource state, $\M{\Xi}_{AB}$, in the satellite. Part $A$ of $\M{\Xi}_{AB}$  is sent through the atmosphere to the ground station, where it is combined with the input state using a balanced beam-splitter. Afterwards, a Bell projective measurement (using a pair of homodyne detectors) on part $A$ and the input state is performed. The measurement result is broadcast to the satellite which, by doing a corrective operation on $B$, recovers the input state as the final output of the protocol.\footnote{We assume noiseless classical communications between satellite and ground station by means of a different channel, such as radio-waves or wide-beam optical signals.}
To describe the teleportation protocol we follow the methodology introduced
in \cite{MarianMarian}. Using this methodology the output state can be computed by using the Wigner characteristic functions (CF) of the input state $\M{\Xi}_\mathrm{in}$ and the entangled resource state $\M{\Xi}_{AB}$. Here, we indicate CFs by $\bigchi(\xi)$, for some complex parameter $\xi$.
In \cite{DellAnno1} the methodology is further expanded to include imperfect homodyne measurements, obtaining
\begin{align}
  \bigchi_{\mathrm{out}}(\xi) =& \bigchi_{\mathrm{in}}(g \eta \xi) \nonumber \\
  &\times \bigchi_{AB}(\xi, g \eta  \xi^*) e^{-\frac{|\xi|^2}{2} g^2 (1 - \eta^2) },
   \label{eq:marian}
\end{align}
where $g$ is the gain parameter, and $\eta$ the efficiency of the homodyne measurements. The CF of a generic $n$-mode state $\M{\Xi}$ is obtained by taking the trace of the product of $\M{\Xi}$ with the displacement operator, giving
\begin{align}
  \bigchi(\xi_1, \xi_2, ..., \xi_n) = \tr\{\M{\Xi} \M{D}(\xi_1)\M{D}(\xi_2) \cdots \M{D}(\xi_n) \},
\label{eq:def_cfunc}
\end{align}
where $\{\xi_1, \xi_2, ..., \xi_n\} \in \mathbb{C}$ are complex arguments, each one representing a mode of $\M{\Xi}$ in the CF.

In this work, we consider that the entangled resource used is a TMSV state.
The TMSV state can be considered as the application of the two mode squeezing operator to the vacuum
\begin{align}
  \ket{\mathrm{TMSV}} = S_{12}(\varrho) \ket{0,0} = e^{\varrho \M{A}_1 \M{A}_2 - \varrho^* \M{A}^\dagger_1 \M{A}^\dagger_2}\ket{0,0},
\end{align}
where $\varrho = r e^{i \phi}$ is the squeezing parameter. Here, we will take $\phi=0$, for simplicity. The CF of a TMSV state is
\begin{align}
\bigchi_\mathrm{TMSV}(\xi_{\mathrm{A}}, \xi_{\mathrm{B}}) = \exp \Big[ & -\frac{1}{2}\Big(V (|\xi_{\mathrm{A}}|^2 + |\xi_{\mathrm{B}}|^2 ) \nonumber \\
&+ \sqrt{V^2 -1}(\xi_{\mathrm{A}} \xi_{\mathrm{B}} + \xi_{\mathrm{A}}^* \xi_{\mathrm{B}}^*) \Big) \Big],
\label{eq:CF_TMSV}
\end{align}
where $V = \cosh(2r)$ is the variance of the distribution of the quadratures. Throughout this work quadrature variances are in shot noise units (SNU), where the variance of the vacuum state is 1 SNU ($\hbar=2$).
Additionally, a coherent state (the state we wish to transfer to the satellite) can considered as the application of the displacement operator to the vacuum,
\begin{align}
  \ket{\alpha} = \M{D}(\alpha)\ket{0},
\end{align}
with the corresponding CF given by
\begin{align}
  \bigchi_{\ket{\alpha}} (\xi) = e^{-\frac{1}{2}|\xi|^2 + 2i\text{Im}(\xi \alpha^*)}.
  \label{eq:CF_coherent}
\end{align}

In general, we can describe the effects  a noisy channel, with a given transmissivity $T$ and excess noise $\epsilon$, has on a mode of any quantum state by scaling the $\xi's$ in the relevant CF by $\sqrt{T}$, and adding a CF corresponding to a vacuum state. For a TMSV state where only mode $B$ is transmitted through the noisy channel, the corresponding CF is \cite{li2011time},
\begin{align}
\bigchi'_\mathrm{TMSV}(\xi_{\mathrm{A}}, \xi_{\mathrm{B}})&=\exp{\left[-\frac{1}{2}(\epsilon + 1 - T)|\xi_{\mathrm{B}}|^2\right]} \nonumber  \\
& \times ~\bigchi_\mathrm{TMSV}(\xi_{\mathrm{A}}, \sqrt{T} \xi_{\mathrm{B}}).
\label{eq:CF_loss}
\end{align}
At times, it will be convenient to refer to the transmissivity in dB, as given by $- 10 \log_{10} T$. Note, that due to the negative sign in this definition, when the transmissivity is referred to in dB a larger loss will have a higher numerical value of the dB transmissivity. Indeed, in this work take the term ``loss'' to mean a transmissivity given in dB - the specific transmissivity being referred to being clear given the context.
If transmissivity is specified without reference to units then it has its normal meaning of a ratio of energies (larger loss corresponding to lower transmissivity).

\subsection{Fidelity of teleportation}
We will use the fidelity as the figure-of-merit to evaluate the effectiveness of quantum teleportation. The fidelity, $F$, is a measurement of the closeness of two states $\M{\Xi}_1$ and $\M{\Xi}_2$, and is given by
\begin{align}
\mathcal{F} = \frac{1}{\pi} \int d^2 \xi \bigchi_{\M{\Xi}_1}(\xi) \bigchi_{\M{\Xi}_2}(-\xi).
\label{eq:fidelity}
\end{align}

To compute the fidelity of a teleported coherent state, $\mathcal{F}^\mathrm{T}$, we first use Eq. (\ref{eq:CF_TMSV}), and Eq. (\ref{eq:CF_loss})
to write the CF of a TMSV state that has been transmitted through a noisy channel. Thereafter, using Eq. (\ref{eq:marian}), and Eq. (\ref{eq:CF_coherent}) we obtain the CF of the teleported state. Finally, $\mathcal{F}^\mathrm{T}$ is  computed as in Eq. (\ref{eq:fidelity}), resulting in
\begin{align}
\mathcal{F}^\mathrm{T}(V, T, \epsilon, \eta, g, \alpha) = \frac{2}{\Delta} \exp \left[-\frac{2}{\Delta} |\alpha|^2 (1- \tilde{g})^2 \right],
\end{align}
where $\tilde{g}=g \eta$, and
\begin{align}
 \Delta = V + \tilde{g}^2 T(V-1)  + \tilde{g}^2 (\epsilon + 1) - 2 \tilde{g}\sqrt{T(V^2-1)} \nonumber \\
         + \tilde{g} + 1 + g^2(1-\eta^2).
\end{align}
Ultimately, $\mathcal{F}^\mathrm{T}$ depends on the characteristics of the noisy channel involved in the protocol ($T$ and $\epsilon$), the parameter $V$, and the gain $g$. These last two parameters, $V$ and $g$, can be controlled to optimize the fidelity teleportation for any given $T$ and $\epsilon$.

We will compare the resulting fidelity of the teleported states with the fidelity of states directly transmitted through the uplink noisy channel. The fidelity of direct transmission, $\mathcal{F}^\mathrm{DT}$, is computed by first writing the CF of a coherent state that has been transmitted through the noisy channel, as
\begin{align}
\bigchi'_{ \ket{\alpha} }(\xi) = \exp{\left[-\frac{1}{2}(\epsilon + 1 - T)|\xi|^2\right]}\bigchi_{ \ket{\alpha} } (\sqrt{T} \xi).
\end{align}
Thereafter, the fidelity between the original state and the transmitted one is computed by using Eq. (\ref{eq:fidelity}), resulting in
\begin{align}\label{eq:dtcoherent}
  \mathcal{F}^\mathrm{DT}(V, T, \epsilon, \alpha ) = \frac{2}{2 + \epsilon} \exp\Big[-\frac{2 (1 - \sqrt{T})^2|\alpha|^2}{2+ \epsilon}\Big].
\end{align}

\begin{figure}
\centering
\includegraphics[width=.48\textwidth]{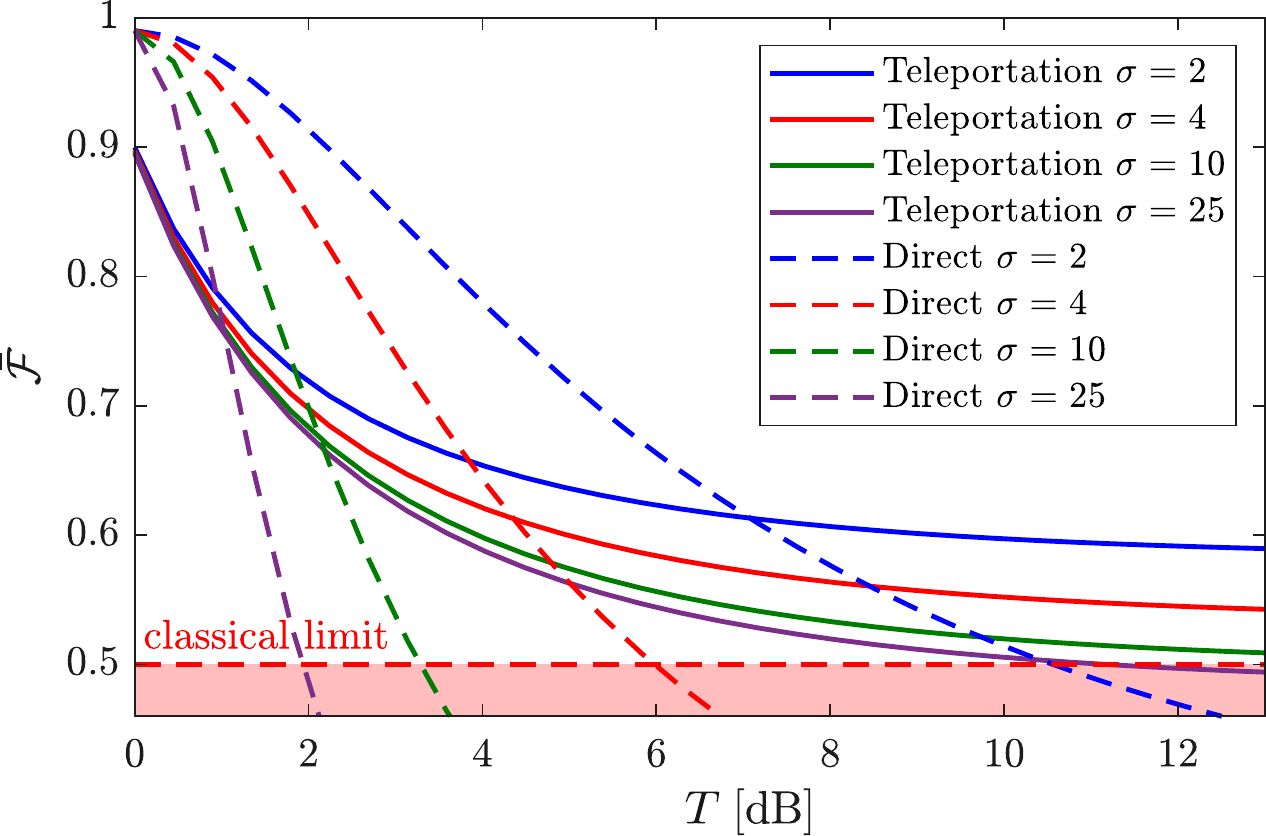}
\caption{Fidelities for teleportation and direct transmission towards the satellite, via a fixed-transmissivity channel. Ensembles of coherent states with different values of $\sigma$ are considered. The shaded area in red marks the region where the teleportation fidelity falls below that achievable using classical communications only. Recall, a higher $T$ in dB corresponds to higher loss.}
\label{fig:fixed}
\end{figure}

To perform a fair assessment, it is not enough to simply consider a single coherent state. Instead, we must consider the mean fidelity over an ensemble of coherent states, drawn from a Gaussian distribution, whose probability distribution is given by \cite{DellAnno1}
\begin{align}\label{eq:probcoherent}
P(\alpha) = \frac{1}{\sigma \pi} \exp\Big(-\frac{|\alpha|^2}{\sigma}\Big),
\end{align}
with $\sigma$ the variance of the distribution. We can think of $\sigma$ as determining the alphabet of states used when transmitting quantum information, or during a protocol such as CV-QKD.
We can now define the mean fidelity as
\begin{align}\label{eq:avefidelity}
\bar{\mathcal{F}}  = \int d\alpha^2 P(\alpha) \mathcal{F}(...,\alpha).
\end{align}
To compare the effectiveness of teleportation relative to direct transmission, we present in Fig. \ref{fig:fixed} the values of $\bar{\mathcal{F}}$ obtained for transmission via a fixed noisy channel, for different values of $\sigma$. The excess noise in the channel is fixed as $\epsilon=0.02$. Throughout this work the efficiency of the homodyne measurements involved in the teleportation is fixed to $\eta=0.9$. Additionally, the values of $g$ and $V$ involved in the teleportation are optimized for each value of the transmissivity. When the loss is small (1 dB), the optimal value of $V$ is approximately 100, however, as the
loss of the channel increases the optimal value of $V$ rapidly decreases towards unity. Using purely classical communications, a value of $F_\mathrm{classical}=0.5$ can be achieved, therefore quantum state transfer is only of interest in the regime where $F>F_\mathrm{classical}$ \cite{Pirandola2006}. 
From the results presented in Fig. \ref{fig:fixed}, we make two observations: First, for each value of $\sigma$, there exists a threshold in the transmissivity above which teleportation yields a higher mean fidelity. Second, as $\sigma$ increases this threshold decreases. This second observation is important for numerous quantum communications protocols (e.g. coherent CV-QKD) in which the more  states that can be transmitted, the better. These two observations indicate that the transmission of quantum states by means of teleportation can be a better alternative relative to simple direct transmission. In the next section we will explore this result in more detail in the context of uplink satellite communications, where we consider teleportation from the ground station to the satellite via a TMSV state created via the downlink channel.

\section{Ground-to-satellite state quantum communication}
We consider a quantum communications setup between a ground station and a satellite.
In this setup, the satellite and ground station have the ability to send and receive quantum optical signals between each other. The ground station is positioned at ground level, $h_0=0$km, and
the satellite when directly overhead at an altitude $H = 500$km. The total propagation length between the satellite and the ground station depends on the zenith angle, $\zeta$, of the satellite relative to the ground station.
%{\color{blue}  For simplicity, we assume that the quantum signals have the same shape independently of their source of transmission, whether it is from the satellite or from the ground station.}
The quantum signals are in the form of short laser pulses with a time-bin width of $\tau_0 = 100$ps, emitted from a laser with a wavelength of $\lambda = 1550$nm. Each laser pulse has an amplitude in the transverse plane possessing a Gaussian profile, and with a beam waist of radius $w_0$. Although in some special configurations the beam $w_0$
can be made as large as the  transmitting aperture, without
loss of generality, we will assume $w_0$  is always smaller than the radius of
the transmitting aperture. As the signal propagates, its beam width increases due to natural diffraction as well as due to the effects of the atmosphere. The satellite and ground station are both equipped with a telescopic aperture to receive the quantum signals. The radius of the aperture of the satellite is $r_\mathrm{sat}$, while for the ground station the radius is $r_\mathrm{gs}$. Besides the quantum signals, the ground station and the satellite also transmit a strong optical signal which can be used as a phase reference for performing homodyne measurements. This strong signal is commonly called a ``local oscillator'' (LO).
In order to study the transmission of quantum signals through the atmosphere, it is key to have a correct model of the effects of the atmospheric turbulence on the propagating beams. Ultimately, this model will allow us to estimate the values $T$ and $\epsilon$ of the uplink and downlink channels.

\subsection{Modeling atmospheric channels}\label{Sec:PhsScrModelling}
The effects of the atmosphere on a propagating beam are modelled using the phase screen model, based in Kolmogorov's theory \cite{Kolmogorov}. The phase screen model is constructed by subdividing the atmosphere into regions of length $\Delta h_i$. For each region the random phase changes induced to the beam by the atmosphere are compressed into a phase screen. The phase screen is then placed at the start of the propagation length, and the rest of the atmosphere is taken to have a constant refractive index. The result at the end of the entire propagation length is a beam that has been deformed mimicking the effects of the turbulent currents in the atmosphere. Thus, this process recreates what a receiver with an intensity detector would observe. Numerically, the beam is represented by a uniform grid of pixels, each one assigned with a complex number, and the propagation is modelled via a Fourier algorithm \cite{villasenor2020atmospheric}.
Since the result of each beam propagation is random, the simulations are run 10,000 times, in order to obtain a correct estimation of the properties of the channel. A detailed description of the numerical methods used can be found elsewhere,  e.g.~\cite{Martin88,SimAP,FourierOptics}.
%{(temporarily added references here, can be further added/changed - commented by Ziqing)}}.

In the phase screen model, the first requirement is a model of the refractive index structure of the atmosphere, $C_n^2$. We use the widely adopted $\mathrm{H-V}_{5/7}$ model \cite{HVmodel}:
\begin{align}
C_n^2(h) &= 0.00594(v/27)^2 (10^{-5}h)^{10} \exp(-h/1000) \\ \nonumber
          &+ 2.7\times10^{-16} \exp(-h/1500) + A \exp(-h/100),
\end{align}
where $h$ is the altitude in meters, $v=21\mathrm{m/s}$ is the rms windspeed, and $A=1.7 \times 10^{-14}\mathrm{m}^{-2/3}$ the nominal value of $C_n^2$ at ground level. In the $\mathrm{H-V}_{5/7}$ model, the main effects of the turbulence are confined to an altitude of 20km, since for higher altitudes the effects are minimal.
Besides the refractive index, we also need the upper-bounds and lower-bounds to the sizes of the turbulent eddies that make up the turbulent atmosphere. The upper-bounds and lower-bounds are the so-called outer-scale and inner-scale, $L_0$ and $l_0$, respectively. Here, we use the empirical Coulman-Vernin profile to model $L_0$ as a function of the altitude $h$ \cite{outer_scale}
\begin{align}\label{Eq:CV_OSProfile}
L_0(h) = \frac{4}{1 + \left(\frac{h-8500}{2500}\right)^2},
\end{align}
and we set the inner-scale to be a some fraction of the outer-scale, specifically, $l_0=\delta L_0$, where $\delta = 0.005$.

With the atmospheric models specified, we now look into how the phase screens are constructed so as to mimic the effects of the turbulence. Each individual phase screen is created
by performing a fast Fourier transform over a uniform square grid of random
complex numbers, sampled from a Gaussian distribution with
zero mean and variance, given by the spectral density function \cite{SimAP}
\begin{align}
\Phi_{\phi}(\kappa) = 0.49 r_0^{-5/3} \frac{\exp(-\kappa^2/\kappa^2_m)}{{(\kappa^2 + \kappa_0^2)}^{11/6}},
\label{pdf}
\end{align}
where $\kappa$ is the radial spatial frequency on a plane orthogonal to the propagation direction, $\kappa_m = 5.92/l_0$, $\kappa_0= 2\pi/L_0$, and $r_0$ is the coherent length. Since the main effects induced by the atmosphere  happen between zero altitude and 20km, the uplink and downlink transmissions will possess key differences, mainly arising from the interplay between the sizes of the beam size and the turbulent eddies. During downlink transmission the beam first encounters the atmosphere with a large beam size - possessing essentially no curvature at this point. On the other hand, in the uplink channel the beam encounters the atmosphere at the start of its path where it has a positive curvature and a small beam size. For these reasons, we expect that the loss in the downlink will be dominated by refraction while the (higher) loss in uplink will be dominated by beam wandering.
Under the flat beam assumption, the coherent length for the downlink can be written as
\begin{align}
  r_0^\mathrm{downlink} = \Big( 0.423 k^2 \sec(\zeta)\int_{h^-}^{h^+} C_n^2(h) dh \Big)^{-3/5},
\end{align}
where $k=2\pi/\lambda$, and $h^-$ and $h^+$ correspond to the lower and upper altitudes of the propagation path corresponding to the respective phase screen.
For the uplink, we need to define first some parameters that characterize the properties of the beam, namely
\begin{align}
\Theta = 1 + \frac{L}{R}, ~~~~ \Lambda= \frac{2L}{k w},
\end{align}
where $R$ and $w$,
%are the curvature radius and the beam waist, respectively, given by
are given by
\begin{align}
R = L \Big[ 1 + \Big( \frac{\pi w_0}{\lambda L} \Big) \Big],   ~~~~~ w= w_0 \Big[1 + \Big( \frac{\lambda L }{\pi w_0} \Big) \Big]^{1/2},
\end{align}
where $L$ is the total distance between satellite and ground station (dependent on $\zeta$).
Given these definitions, the coherent length for the uplink channel can be written as
\begin{align}
r_0^\mathrm{uplink} = \Big( 0.424 k^2 \sec(\zeta) (\mu_1 + 0.622 \mu_2 \Lambda^{11/6}) \Big)^{-3/5},
\end{align}
where
\begin{align}
\mu_1 &= \int_{h^-}^{h^+} C_n^2(h) \Big[ \Theta \Big( \frac{H-h}{H - h_0} \Big) + \frac{h-h_0}{H-h_0} \Big]^{5/3}, \\ \nonumber
\mu_2 &= \int_{h^-}^{h^+} C_n^2(h) \Big[ 1 - \frac{h-h_0}{H-h_0} \Big]^{5/3}. \nonumber
\end{align}

\begin{figure}
\centering
\includegraphics[width=.5\textwidth]{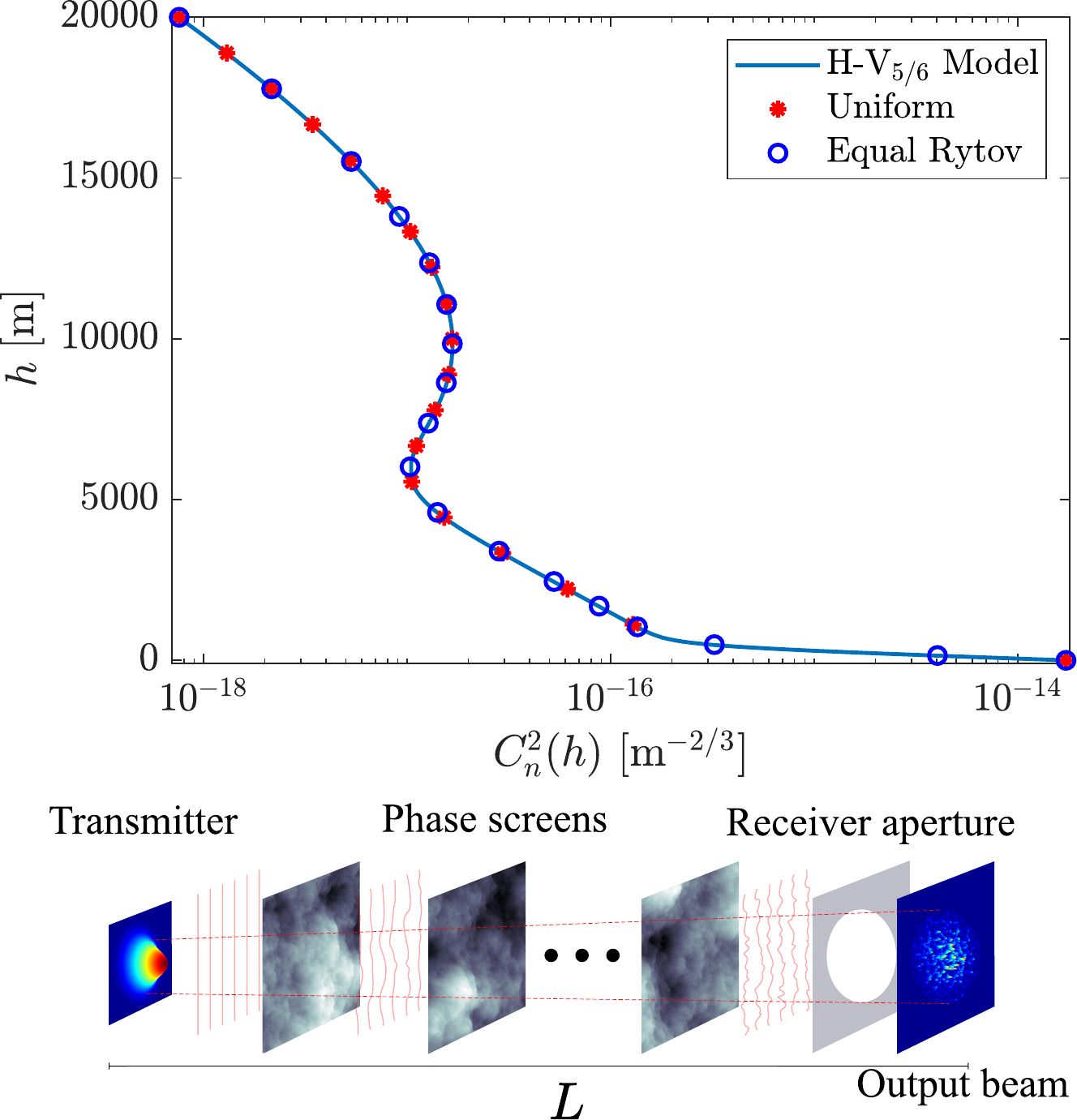}
\caption{The H-V${}_{5/7}$ model, with the positions of the phase screens shown for two cases: equally spaced phase screens and spacing that conserves a constant value of the Rytov parameter with $b=0.2$.}
\label{fig:HVmodel}
\end{figure}

The position of each phase screen is determined using the condition that the Rytov parameter, $r_R^2$, is maintained constant over each length $\Delta h_i$, specifically \cite{8934090}
\begin{align}
r_R^2 = 1.23 k^{7/6} \int_{h^-}^{h^+} C_n^2(h) (h-h^-)^{11/6} dh = b.
\label{eq:Rytov}
\end{align}
We set a value of $b=0.2$, which corresponds to a total of 17 phase screens up to 20km.
In Fig. \ref{fig:HVmodel}, we plot the $\mathrm{H-V}_{5/7}$ model, with the positions of the phase screens set by the condition given by Eq. (\ref{eq:Rytov}). For comparison, we also plot the positions of the phase screens placed at a uniform distance between ground level and 20km. We can see that by using the condition imposed by Eq. (\ref{eq:Rytov}) the phase screens are more adequately distributed to account for the altitude  variations in the turbulence.
Finally, to account for the remaining turbulence between 20km and $H$ a single phase screen is used.

At the end of every beam propagation simulation we can obtain the transmissivity induced by the atmosphere by integrating the intensity of the beam over the receiver aperture, as
\begin{align}
T_\mathrm{turb} = \frac{\iint_{\mathcal{D}} I_\mathrm{sig} dA}{P_0},
\end{align}
where $I_\mathrm{sig}$ is the intensity (power per unit area) of the beam at the plane containing the receiver aperture, $P_0$ is the initial total power of the beam at the point of emission, and $\mathcal{D}$ is the surface area of the receiver aperture.
Despite the main source of loss arising from the atmospheric turbulence, we also need to account for the extinction of the signal caused by absorption and scattering by the particles of the atmosphere, as well as the loss due to the imperfect optical devices used. To account for the extinction we adopt a transmissivity $T_\mathrm{ext}= \exp(-0.7 \sec{\zeta})$. For the loss due to the optical devices we consider a   transmissivity value $T_\mathrm{opt} = 0.794$ (1dB)
\cite{EllipticBeamGeneralized2019}. The total transmissivity of the channel is then simply,
\begin{align}
T = T_\mathrm{turb} T_\mathrm{ext} T_\mathrm{opt}.
\end{align}

\subsection{Excess noise}
Since in CV quantum states the information is encoded in the quadratures of the states, we require an LO in order to extract this information via  homodyne or heterodyne measurements. With this in mind, we can take the results presented in this work which adopt a non-zero $\epsilon$, as the fidelity outcomes expected if we were to actually measure the fidelities
experimentally \cite{TeleportationExperiment}. The ideal theoretical predictions would correspond to the pure loss channel case, where $\epsilon=0$.
In \cite{Sebastian} it is discussed that for coherent state transmission via atmospheric channels the main components of the excess noise arise from turbulence-induced effects on the LO, in addition to time-of-arrival fluctuations caused by delays between the laser pulses and the LO. The variations in the intensity of the LO induce an excess noise given by
\begin{align}
  \epsilon_{ri} = \sigma_{\mathrm{SI,LO}}^2(D) V_\mathrm{sig},
  \label{eq:noise_ri}
\end{align}
where $V_\mathrm{sig}$ the statistical variance of the quadratures of the quantum signal, corresponding to $V_\mathrm{sig}=\sigma$ for direct transmission, and $V_\mathrm{sig}=V$ for the teleportation channel.
For a given aperture size, the scintillation index averaged over the aperture of the LO is
\begin{align}
\sigma_\mathrm{SI,LO}(\mathcal{D})=\langle P_{LO}^2 \rangle / \langle P_{LO} \rangle^2-1,
\end{align}
where $P_{LO} = \iint_{\mathcal{D}} I_\mathrm{LO} dA $ is the power of the LO (with intensity given by $I_\mathrm{LO}$) over the aperture. Since the uplink channel is more affected by beam wandering, $\sigma_{SI,LO}(\mathcal{D})$ can be expected to be much greater  for the uplink relative to the downlink.

Time-of-arrival fluctuations are caused by a broadening of the time-bin width of the signal pulse from $\tau_0$ to $\tau_1$, where $\tau_1$ is given by \cite{Sebastian}
\begin{align}
  \tau_1 = \sqrt{\tau_0^2 + 8 \mu},
\end{align}
where
\begin{align}
  \mu &= \frac{0.391(1 + 0.171 \delta^2 - 0.287\delta^{5/3})\upsilon_1 \sec(\zeta)} {c^2}, \\
  \upsilon_1 &= \int_{h_0}^H C_n^2 L_0^{5/3} dh, \nonumber
\end{align}
and where $c$ is the speed of light in vacuum.
As derived in \cite{Young:98}, the variance of $\tau_1$, is given by $\sigma_{\mathrm{ta}}^2 ={\tau_1^2}/{4}$, which leads to an excess noise \cite{AtmosphericQKD}:
\begin{align}
  \epsilon_\mathrm{ta}=2 (k c)^2 (1-\rho_\mathrm{ta}) \sigma_{\mathrm{ta}}^2 V_\mathrm{sig},
  \label{eq:noise_ta}
\end{align}
where $\rho_\mathrm{ta}$ is the timing correlation coefficient between the LO and the signal. The value of $\sigma_{\mathrm{ta}}$ is independent of the direction of propagation of the beam.
For a value of $\tau_0=100$ps, $\epsilon_\mathrm{ta}$ is virtually independent of the atmospheric turbulence, since the pulse broadening only becomes considerable for $\tau_0 < 0.1$ps \cite{Sebastian}. Therefore, considering that $\rho_\mathrm{ta}= 1 - 10^{-13}$, the noise contribution due to the time of arrival fluctuations becomes $\epsilon_\mathrm{ta}=0.007 V_\mathrm{sig}$.

With the two main sources of noise outlined, we now write the total excess noise as $\epsilon = \epsilon_{\mathrm{ta}}+\epsilon_{\mathrm{ri}}$.
The excess noise being directly proportional to $V_\mathrm{sig}$ reflects the fact that due to the fluctuating nature of atmospheric channels, the values of $T$ and $\epsilon$ need to be estimated by repeated measurements of the channel. This means that in a experimental setup one cannot distinguish between variations of the quadratures due to quantum uncertainty, or the variations induced by the fluctuating value of $T$. Therefore, the variations of $T$ of the channel effectively translate to additional excess noise.
We note that there are additional sources of excess noise, however, their contributions are minor compared to those considered here \cite{PracticalCVQKD}.

\begin{table}
\centering
\caption{Satellite channel parameters.}
\begin{tabular}[t]{ccccccc}
$H$   &
$h_0$ &
$\lambda$ &
$\tau_0$ \\
\hline
500km &
0km &
1550nm &
100ps \\
\hline
\label{tab:params}
\end{tabular}
\vspace{-5mm}
\end{table}

%------------ start of Ziqing's paragraphs--------------
\subsection{Other channel modeling techniques}
Throughout this work we use the phase screen simulations to model the channel.
% to simulate the atmospheric propagation of an optical beam.
Performing phase screen simulations is essentially a numerical approach to solving the \textit{stochastic parabolic equation}, and adopts a versatile technique referred to as the \textit{split-step} method~\cite{SimAP}.
Despite  its computationally intensive nature, the split-step method has been widely used to study the atmospheric optical propagation of classical light under a variety of conditions (see e.g.~\cite{SSM2,SSM3,SSM4,SSM5,SSM6,SSM7}).
Due to its quantitative agreement with analytical results, the split-step method is also believed to be
%capable of simulating the atmospheric propagation of quantum light
very reliable (see e.g.~\cite{OAM_Entanglement_MPS,EntanglementProtectionAO2018,EntanglementProtectionAO2019}).

Other channel-modeling techniques have been proposed to simplify the description of the atmospheric propagation of quantum light under specific situations. It is worthwhile to compare their predictions with our detailed phase screen simulations.
Channel modeling techniques based on the so-called \textit{elliptic-beam approximation}~\cite{AtmosphericChannels} are believed to be particularly useful when the phase fluctuations of the \textit{output} field amplitude can be neglected. This point is discussed further in~\cite{EllipticBeamWeather2017} where it is also highlighted that homodyne measurements can be constructed where phase fluctuations of the output field can be neglected.
%As discussed in e.g.~\cite{EllipticBeamWeather2017}, the channel modeling techniques based on the so-called \textit{elliptic-beam approximation}~\cite{AtmosphericChannels} are believed to be useful when the phase fluctuations of the output field amplitude can be neglected (this is the case in this work due to the use of homodyne measurements).
%We highlight the channel modeling techniques based on the so-called \textit{elliptic-beam approximation}~\cite{AtmosphericChannels} are believed to be useful when the phase fluctuations of the output field amplitude can be neglected (this is the case in this work due to the use of homodyne measurements, see discussions in e.g.~\cite{EllipticBeamWeather2017}).
Under the elliptic-beam approximation, it is assumed that the atmospheric propagation leads to only beam wandering, beam spreading, and beam deformation (into an elliptical form). However, the extinction losses due to back-scattering and absorption can also be added phenomenologically under such an approximation~\cite{EllipticBeamWeather2017}.
Although originally proposed under the assumption of a horizontal channel, the elliptic-beam approximation was directly adopted in~\cite{ChannelParamterEstimation1} to study the performance of CV-QKD in the downlink channel. In addition, the authors of~\cite{EllipticBeamGeneralized2019} proposed a generalized channel modeling technique based on the elliptic-beam approximation, providing a comprehensive model for the losses suffered by the quantum light in both the uplink and downlink channels. All these works~\cite{AtmosphericChannels,EllipticBeamWeather2017, ChannelParamterEstimation1, EllipticBeamGeneralized2019} assumed an infinite outer scale (i.e. $L_0\!=\!\infty$) and a zero inner scale (i.e. $l_0\!=\!0$), effectively neglecting the inner scale and outer scale effects.
%. Despite of the simplicity provided by such an assumption, these works

\begin{figure}[htb!]
	\centering
	\includegraphics[width=1\linewidth]{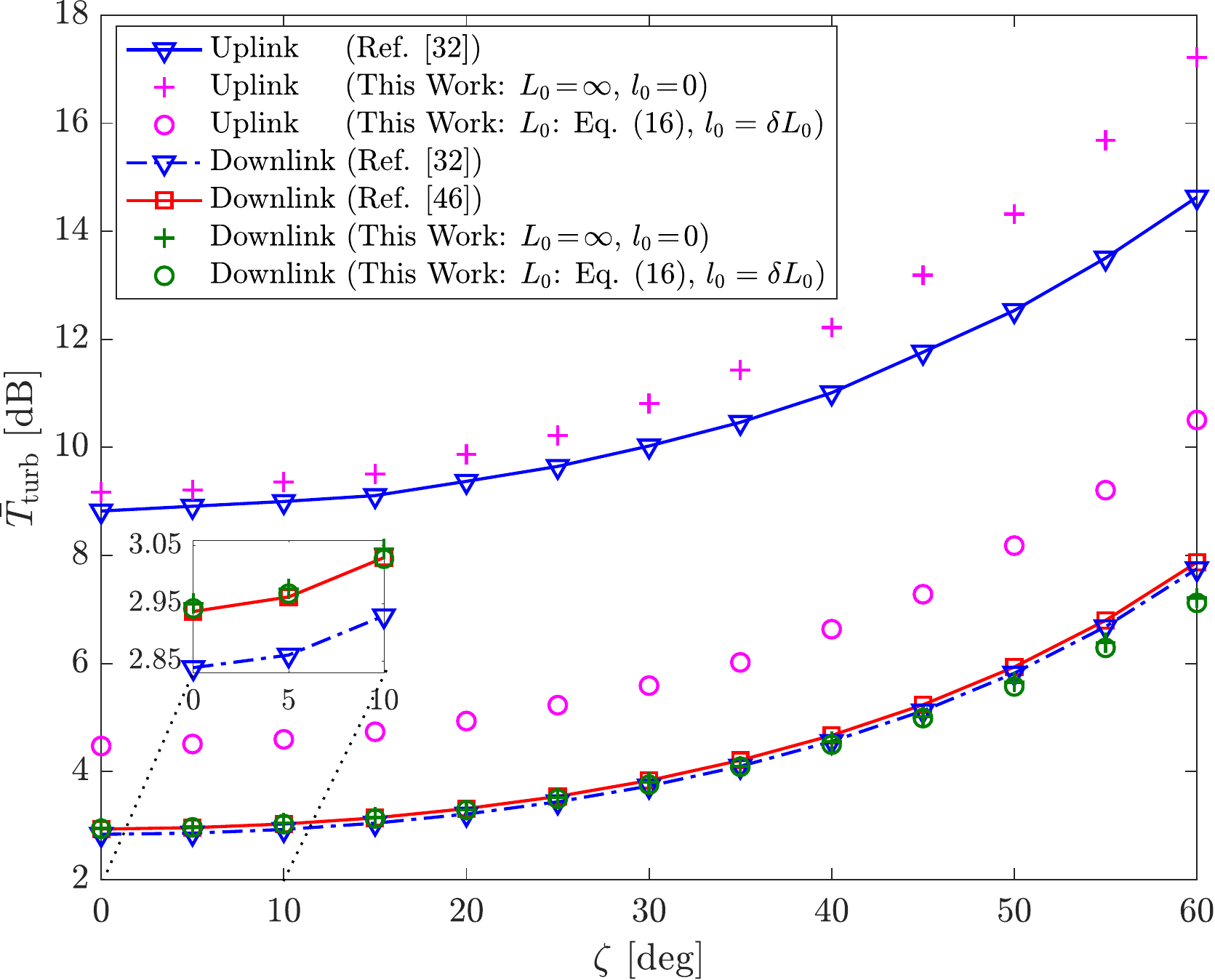}
	\caption{The mean turbulence-induced loss $\bar{T}_{\text{turb}}\,[\text{dB}]$ predicted by different channel modeling techniques.
	By ``this work'' we mean the phase screen simulations we have undertaken.
	The parameters are given in Table~\ref{tab:params}, with $w_0=15$ cm and $r_\mathrm{sat}=r_\mathrm{gs}=1$ m. Recall, a higher $\bar{T}_{\text{turb}}$ in dB corresponds to higher loss.}
%		Note, that only the turbulence-induced losses are compared. Note again that this figure is for $w_0=15$cm, and $r_\mathrm{sat}=r_\mathrm{gs}=1\,\mathrm{m}$.

	\label{fig:fig_T_CMP}
\end{figure}

In Fig.~\ref{fig:fig_T_CMP}, we compare the {mean turbulence-induced loss $\bar{T}_{\text{turb}}\,[\text{dB}]$} predicted by i)~the phase screen simulations, and ii)~the channel modeling techniques (based on the elliptic-beam approximation) of~\cite{ChannelParamterEstimation1} and~\cite{EllipticBeamGeneralized2019}.
%Note, that only the turbulence-induced losses are compared.
Although our phase screen simulations take into account the inner scale and outer scale effects by adopting the empirical Coulman-Vernin profile (recall Eq.~(\ref{Eq:CV_OSProfile})), for comparison we also present the results predicted by the phase screen simulations with $L_0\!=\!\infty$ and $l_0\!=\!0$.
From Fig.~\ref{fig:fig_T_CMP} we clearly observe that the mean transmissivity in the downlink channel predicted by all the considered channel modeling techniques are similar. This can be explained by the fact that the main source of loss in a downlink channel is diffraction loss.
For the uplink channel, we observe that the mean transmissivities predicted by the phase screen simulations with $L_0\!=\!\infty$ and $l_0\!=\!0$ match the mean transmissivities predicted by the generalized channel modeling technique.
% (based on the elliptic-beam approximation) proposed in~\cite{EllipticBeamGeneralized2019} reasonably well.
Such an observation is reasonable since~\cite{EllipticBeamGeneralized2019} indeed assumes $L_0\!=\!\infty$ and $l_0\!=\!0$.

An interesting observation from Fig.~\ref{fig:fig_T_CMP} is that the mean losses predicted with a finite outer scale and a non-zero inner scale are lower than the mean losses predicted with an infinite outer scale and a zero inner scale. Such an observation  can be explained mainly by the fact that the presence of a finite outer scale reduces the amount of beam wandering and long-term beam spreading~\cite{andrews_book1}. This observation does not refute the conventional wisdom that the channel loss in the uplink channel is higher than the channel loss in the downlink channel. However, this observation does indicate that the disadvantage of an uplink channel may be overestimated in some models.
We believe that setting a finite outer scale and a non-zero inner scale (according to the empirical Coulman-Vernin profile) is more relevant (rather than simply setting $L_0=\infty$ and $l_0=0$) when studying the atmospheric propagation of light through a satellite-based channel. Therefore, in the rest of this work, we will utilize the results from the phase screen simulations that adopted  a finite outer scale and a non-zero inner scale.
%------------ end of Ziqing's paragraphs--------------

\subsection{Ground-to-satellite state transmission}
\begin{figure}
\centering
\includegraphics[width=.48\textwidth]{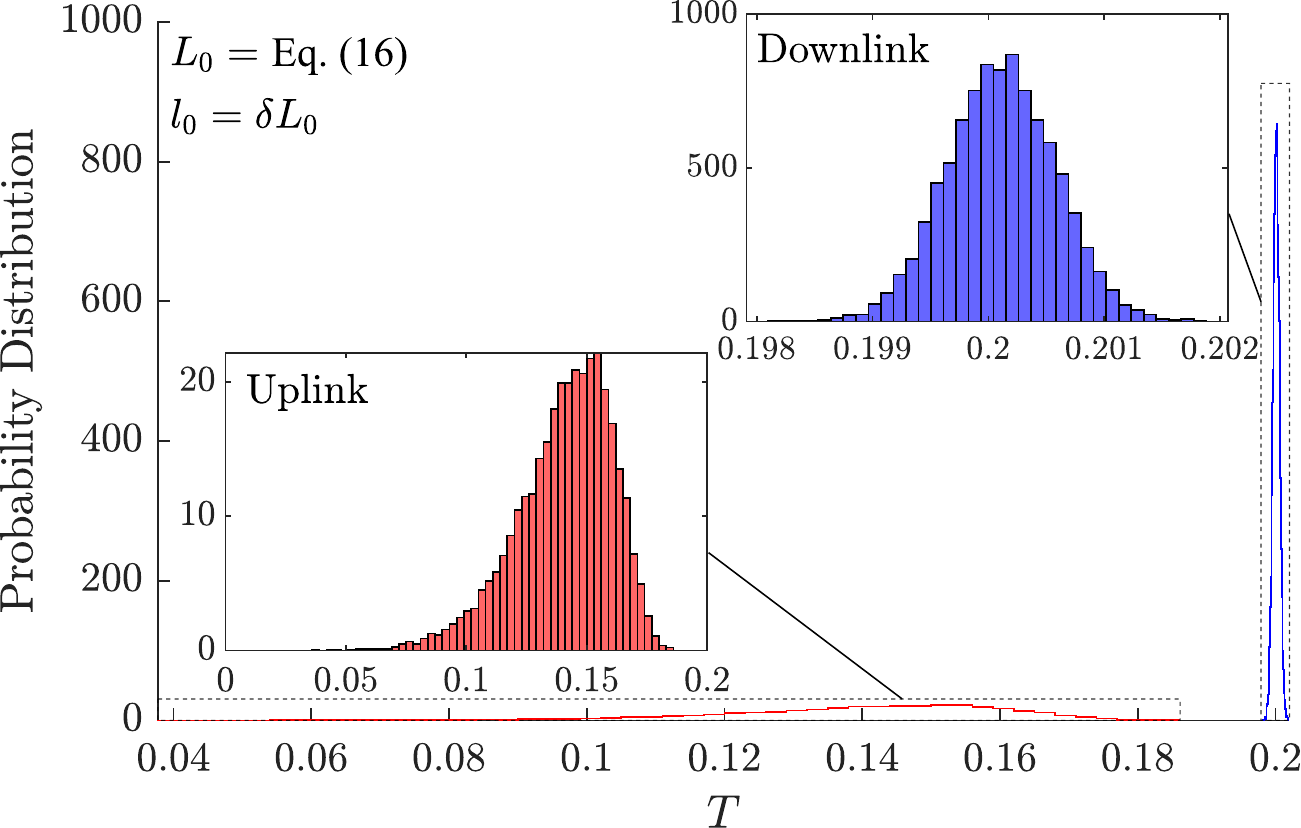}
\caption{Probability distribution functions of the transmissivity for the satellite communications channels, uplink (red) and downlink (blue). The parameters of the channels are given in Table \ref{tab:params}, with $\zeta=0^o$, $w_0=15$cm, and $r_\mathrm{sat}=r_\mathrm{gs}=1$m.}
\label{fig:hist}
\end{figure}

Using our phase screen simulations we model an uplink and a downlink channel with the characteristics presented in Table~\ref{tab:params}. We consider that $r_\mathrm{sat} = r_\mathrm{gs}$ in order to focus our analysis in the turbulence induced loss. We do note, that in a realistic satellite communications deployment it is expected that the aperture of the ground station is larger than the satellite's aperture (see later calculations). However, setting the apertures constant in the first instance allows for a more  direct comparison of the effects of turbulence on the links. The model returns the probability distribution function (PDF) of the loss for each channel, as seen in Fig.~\ref{fig:hist}. The PDF of the downlink channel is extremely narrow compared to the PDF corresponding to the uplink channel. This is due to the asymmetry of the interaction between the beam and the atmosphere, as explained above.
%The figure shows that the mean values of transmissivity for the uplink channel are considerably larger compared to the downlink channel. Moreover, the variance in the distributions of $T$ is far greater for the uplink channel, as a consequence of the presence of beam wandering.
%Additionally, as discussed above, we observe a large disagreement from the uplink model when $L_0$ and $l_0$ are modelled by the Coulman-Vernin model, and when $L_0\!=\!\infty$ and $l_0\!=\!0$.
The scintillation index of the LO is computed by simulating the propagation of a strong beam corresponding to the LO. The scintillation index values are several orders of magnitude larger for the uplink relative to the downlink.

\begin{figure}
\centering
\includegraphics[width=.48\textwidth]{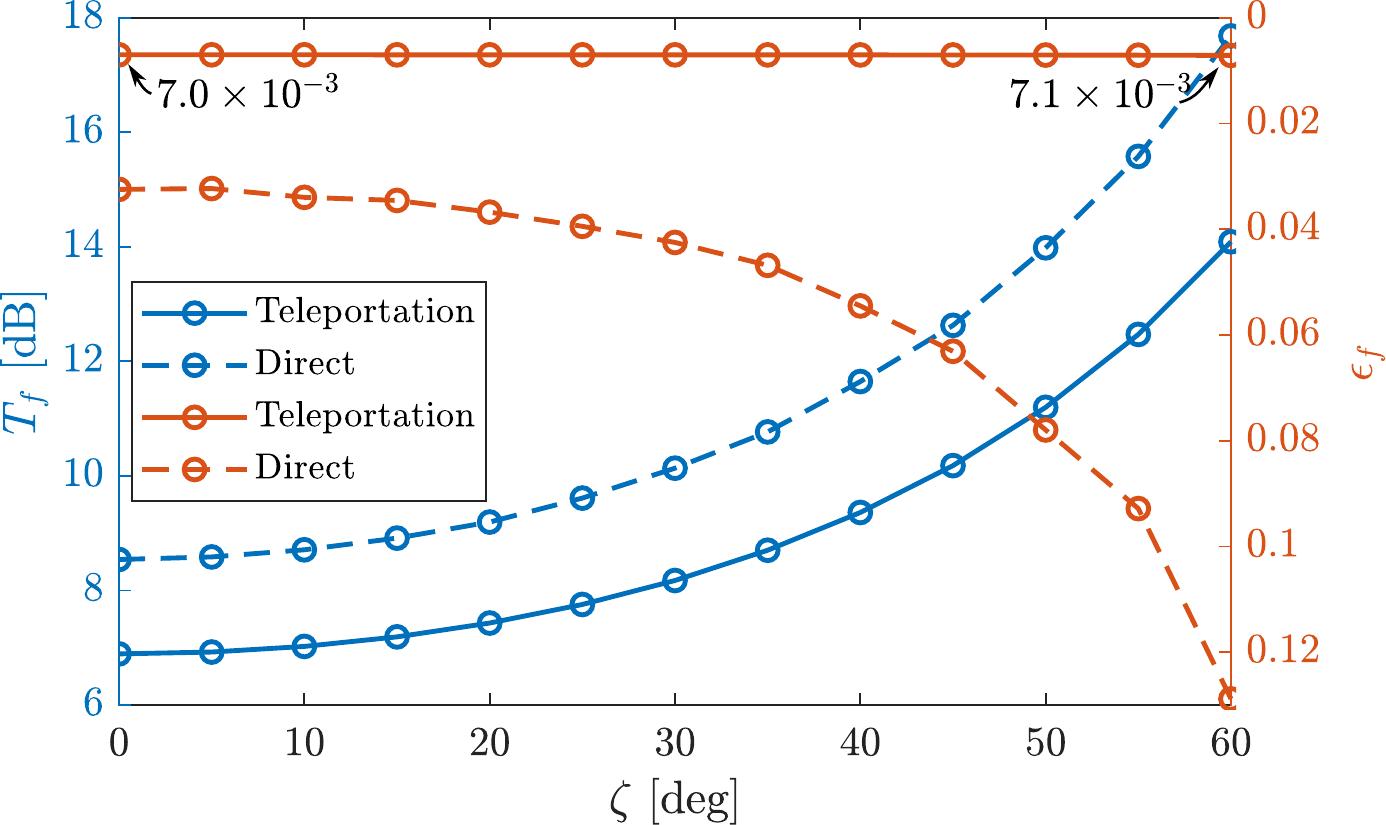}
\caption{Ground-to-satellite  properties for the direct transfer channel and for the teleportation channel, shown for $V_\mathrm{sig}=1$. The parameters of the channels are given in Table \ref{tab:params}, with $w_0=15$cm and $r_\mathrm{sat}=r_\mathrm{gs}=1$m. For the teleportation channel the entangled resource is distributed via the
downlink. The left axis (blue) corresponds to the effective transmissivity, while the right axis (red) corresponds to the effective excess noise. Recall, a higher ${T}_f$ in dB corresponds to higher loss. }
\label{fig:channel}
\end{figure}

Due to the fluctuating nature of the uplink and downlink channels we need to consider ensemble-averages when computing the fidelity of the transmitted states \cite{NedaComposable}. The required analysis can be derived as in the non-fluctuating channel if we define an {\it effective transmissivity} $T_{\mathrm{f}}$, and an {\it effective excess noise} $\epsilon_{\mathrm{f}}$, as
\begin{align}
  &T_{\mathrm{f}} = \overbar{\sqrt{T}}^2, ~~~~~~~
  \epsilon_{\mathrm{f}} = \frac{\mathrm{Var}(\sqrt{T})}{T_{\mathrm{f}}} V_\mathrm{sig} + \epsilon \bar{T}, \\ \nonumber
  &\mathrm{Var}(\sqrt{T}) = \bar{T}  - \overbar{\sqrt{T}}^2,
\end{align}
with the mean values computed as
\begin{align}
  &\bar{T} = \int_0^1 T p_\zeta(T) dT,  ~~~~~
  \overbar{\sqrt{T}} = \int_0^1 \sqrt{T} p_\zeta(T) dT,
\end{align}
with $p_\zeta(T)$ the PDF of $T$ for a given $\zeta$.
We present in Fig. \ref{fig:channel} the properties of the downlink and uplink channels obtained using the phase screen simulations.
Following Eqs. (\ref{eq:noise_ri}, \ref{eq:noise_ta}), the value of $\epsilon_{\mathrm{f}}$ is proportional to the variance of the quadratures of the quantum states transmitted through the channel. For this reason we show on the plot the value of $\epsilon_{\mathrm{f}}$ with a fixed $V_\mathrm{sig}=1$, to give a fair comparison between the two channels, but we emphasize that this parameter will change in our calculations below. We observe that, as expected,  losses are higher (i.e. larger effective transmissivity when stated in dB) for direct transmission. Moreover, the value of $\epsilon_{\mathrm{f}}$ for the direct channel is one order of magnitude greater than the value for the teleportation channel. This is a direct consequence of the variations in the intensity for both the quantum signal and the LO. We do not show the results for direct transmission modelled for an uplink with  $L_0\!=\!\infty$ and $l_0\!=\!0$, but we find that $\epsilon_{\mathrm{f}} \approx 0.6$ for $\zeta=0^o$, meaning such a channel is inadequate for the transmission of quantum states.

\begin{figure}
\centering
\includegraphics[width=.48\textwidth]{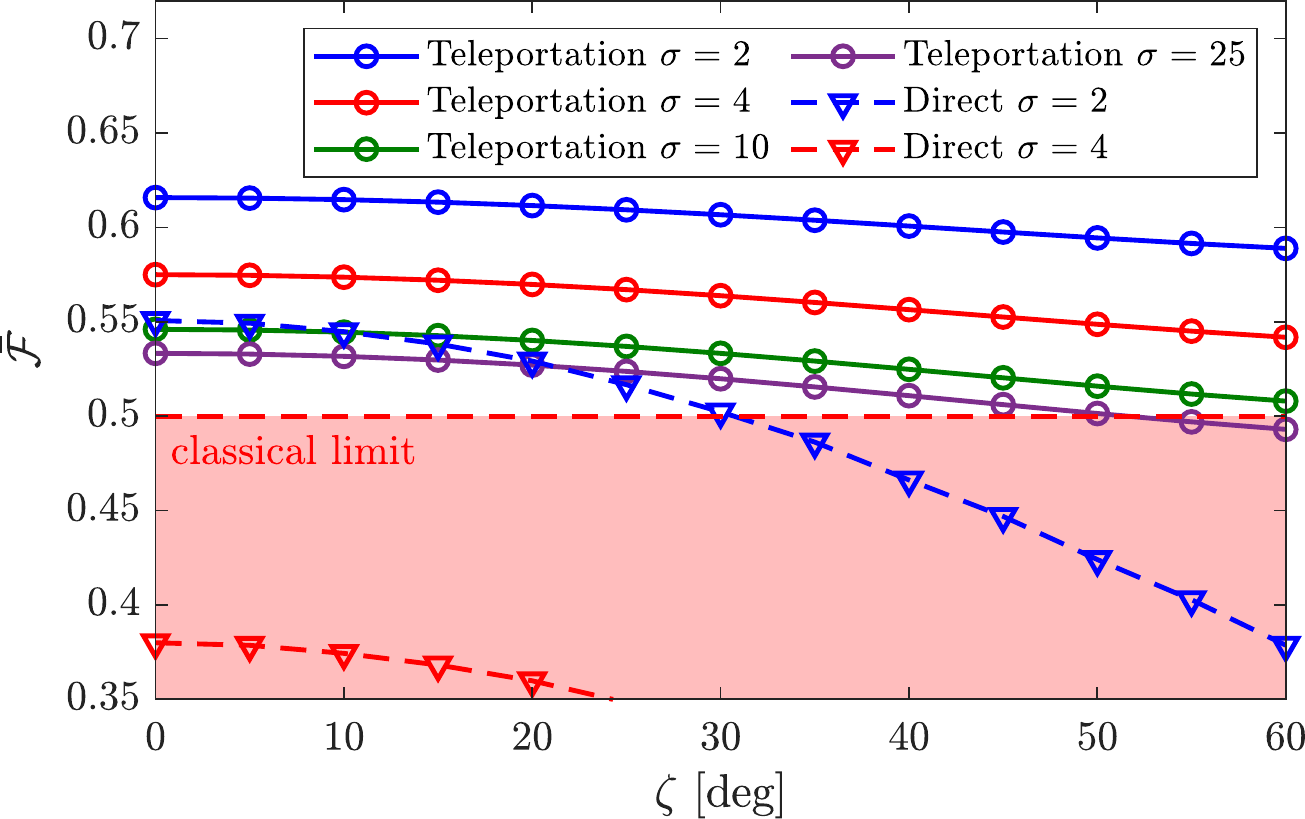}
\caption{Mean fidelities for ground-to-satellite transfer via direct transmission and via teleportation, shown for different values of $\sigma$. The channels parameters adopted are given in Table \ref{tab:params}, with $w_0=15$cm and $r_\mathrm{sat}=r_\mathrm{gs}=1$m. The shaded area in red indicates the region where the fidelity can be achieved by classical communications only. The direct transmission for $\sigma=10, 25$ result in mean fidelities $<0.35$ for all zenith angles.}
\label{fig:fidelities}
\end{figure}

Using the values of $T_{\mathrm{f}}$ and $\epsilon_{\mathrm{f}}$, obtained from the numerical simulations, we then compute the fidelity of teleportation and direct transmission of coherent states. The values of $g$ and $V$ are optimized relative to the loss of the teleportation channel to maximize the mean fidelity.
For the loss values anticipated for the teleportation channel, we observe that the
optimal value of V is in the range 1 to 1.5, and the optimal value of g is in
the range 1 to 1.1.
The results, presented in Fig. \ref{fig:fidelities}, show that the teleportation channel has a significant advantage over direct transmission. We see that direct transmission is only capable of overcoming the classical limit for a reduced alphabet of $\sigma=2$, and low zenith angles up to $30^o$. On the other hand, the teleportation channel exceeds the classical limit for a larger range in the alphabet, and for a wide range of zenith angles. This shows that one can indeed avoid, to a significant extent, the detrimental effects of the direct uplink channel via a teleportation using  an entangled resource distributed via the downlink channel. We note that the values of $\sigma$ considered here encompass the ranges required to undertake high-throughput CV-QKD \cite{PracticalCVQKD}.

\subsubsection{Asymmetric apertures}
A stated earlier, in the calculations just described we assumed that $r_\mathrm{gs}=r_\mathrm{sat}$, in order to focus our analysis in the turbulence induced loss. However, in many satellite deployments it is expected that $r_\mathrm{sat} < r_\mathrm{gs}$. In such a case, use of the teleportation channel in the manner we have described would present an even higher advantage over the direct transmission channel.
For example, we find that for a space communications setting as in Table \ref{tab:params}, with $\zeta=0^o$, $w_0=15$cm and $r_\mathrm{gs}=50$cm, the downlink channel incurs a loss of $\approx 11$dB. Meanwhile, under the same values, but with $r_\mathrm{sat}=20$cm, the uplink channel incurs a higher loss of $\approx 22$dB. This means that the fidelity obtained using the teleportation channel in this setting is approximately 0.6, while the fidelity via direct transmission would be well below the classical limit. It is therefore important to emphasize, that our detailed calculations most likely represent a lower bound on the actual gain in  uplink communications of many future satellite missions.

%%%%%%%%%%%%%%%%%%%%%%%%%%%%%%%%% MINGJIAN's part %%%%%%%%%%%%%%%%%%%%%%%%%%%%%%%%%%%%%%%%%%%%%%%%%%%%%%%%%%%%%%%%%%%%%%%%%%%%
%%%%%%%%%%%%%%%%%%%%%%%%%%%%%%%%%%%%%%%%%%%%%%%%%%%%%%%%%%%%%%%%%%%%%%%%%%%%%%%%%%%%%%%%%%%%%%%%%%%%%%%%%%%%%%%%%%%%%%%%

\section{CV Teleportation with non-Gaussian Operations}
A great deal of recent research has been focused on the photonic engineering of highly non-classical, non-Gaussian states of light, aiming to achieve enhanced entanglement and other desirable properties.
Indeed, non-Gaussian features are essential for various quantum information tasks, such as
%quantum teleportation
%\cite{cochrane2002teleportation,dell2007continuous,yang2009entanglement,dellanno2010realistic,wang2015continuous-variable,xu2015enhancing,hu2017continuous-variable}, quantum key distribution \cite{huang2013performance,borelli2016quantum,li2016non-gaussian,guo2017performance,zhao2017improvement,ma2018continuous-variable,guo2019continuous-variable,PhysRevA.102.012608,he2019multi},
entanglement distillation \cite{lund2009continuous-variable,zhang2010distillation,fiurasek2010distillation,datta2012compact,lee2013entanglement,zhang2013continuous-variable-entanglement,seshadreesan2019continuous-variable,mardani2020continuous},
noiseless linear amplification \cite{ralph2009nondeterministic,kim2012quantum,yang2012continuous-variable,gagatsos2014heralded,zhang2018photon,adnane2019quantum},
and quantum computation \cite{ghose2007non,marshall2015repeat,miyata2016implementation,marshall2016continuous}.
In entanglement distillation and noiseless linear amplification, non-Gaussian features are a requirement due to the impossibility of distilling (or amplifying) entanglement in a pure Gaussian setting \cite{eisert2002distilling}.
In universal quantum computation, non-Gaussian features are indispensable if quantum computational advantages are to be obtained \cite{rahimikeshari2016sufficient}.

Non-Gaussian operations, which map Gaussian states into non-Gaussian states, are a common approach to delivering non-Gaussian features into a quantum system.
At the core of non-Gaussian operations is the application of the annihilation operator $\M{A}$ and the creation operator $\M{A}^\dagger$.
There are two basic types of these operations, namely photon subtraction (PS) and photon addition (PA), which apply $\M{A}$ and $\M{A}^\dagger$ to a state, respectively.
Both operations have been shown to enhance the entanglement of TMSV states (e.g., \cite{navarrete2012enhancing,bartley2013strategies,bartley2015directly}).
Various studies on  combinations of PS and PA have also been undertaken (e.g., \cite{parigi2007probing, yang2009nonclassicality, lee2011enhancing, park2012enhanced}).
A specific combination, photon catalysis (PC), is of particular research interest.
Instead of subtracting or adding photons, PC \textit{replaces} photons from a state, and is known to significantly enhance the entanglement of TMSV states under certain conditions (e.g., \cite{ulanov2015undoing,he2020global}).
If TMSV states are in fact shared between a satellite and a ground station, it is natural to ask whether non-Gaussian operations can be adopted at the ground station to further facilitate satellite-based quantum teleportation.

\subsection{Non-Gaussian states and non-Gaussian operations}
\begin{figure}
	\centering
	\includegraphics[width=1\linewidth]{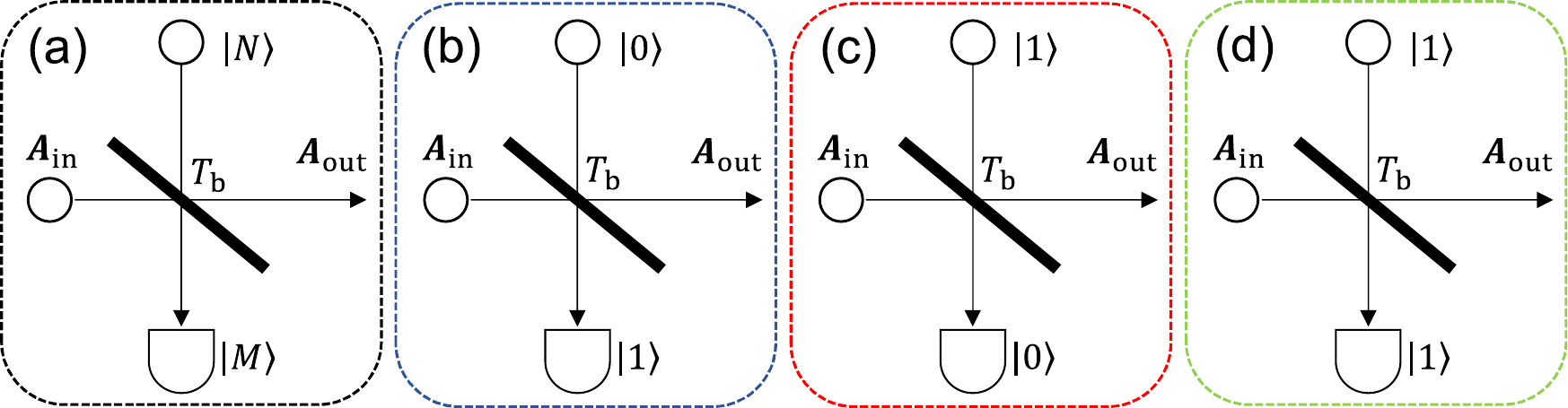}
	\caption{The experimental setups for (a) a wide class of non-Gaussian operations, (b) photon subtraction, (c) photon addition and (d) photon catalysis.}
	\label{fig:diagnongaussian}
\end{figure}

A simple experimental setup for realizing non-Gaussian operations consists of beam-splitters and photon-number-detectors.
For example, as is depicted in Fig.~\ref{fig:diagnongaussian}a, an input state interacts with an ancilla Fock state $\ket{N}$ at a beam-splitter with transmissivity $T_{\mathrm{b}}$.
If $M$ photons are detected in the ancilla output the operation has succeeded.
In practice, the probability of success of a non-Gaussian operation is an important parameter to consider.
In this regard, single-photon non-Gaussian operations ($M,N\in\left\lbrace 0,1 \right\rbrace$) usually have the highest success probability for a given type of non-Gaussian operation \cite{bartley2015directly}, making them the best candidates for practical implementation.
Therefore, in this work we will restrict ourselves to the non-Gaussian operations with single-photon ancillae and single-photon detection (i.e., Fig.~\ref{fig:diagnongaussian}b, c, and d).

In the Schrodinger picture, the transformation of the non-Gaussian operations described above can be represented by an operator \cite{jia2014decompositions}
\begin{equation}
	\M{O}=\bra{M}\M{U}(T_{\mathrm{b}})\ket{N},
\end{equation}
where
\begin{equation}
	\begin{aligned}
		\M{U}(T_{\mathrm{b}})=: \exp \Big\{(\sqrt{T_{\mathrm{b}}}-1)&\left(\M{A}^{\dagger} \M{A}+\M{B}^{\dagger} \M{B}\right) \\
		+&\left(\M{A} \M{B}^{\dagger}-\M{A}^{\dagger} \M{B}\right) \sqrt{(1-T_{\mathrm{b}})}\Big\}:,
	\end{aligned}
\end{equation}
is the beam-splitter operator, $: \cdot :$ means simple ordering (i.e., normal ordering of the creation operators to the left without taking into account the commutation relations), and $\M{A}$ and $\M{B}$ are the annihilation operators of the incoming state and the ancilla, respectively. Using the coherent state representation of the Fock state,
\begin{equation}
	\ket{N}=
	\left.\frac{1}{\sqrt{N !}} \frac{\partial^{N}}{\partial \alpha^{N}}
	\exp \left(\alpha \M{B}^{\dagger}\right)|0\rangle
	\right|_{\alpha=0},
\end{equation}
we can obtain the following compact forms for the operators for PS ($N=0,\,M=1$), PA ($N=1,\,M=0$), and PC ($N=1,\,M=1$) \cite{guo2019continuous-variable}, respectively,
\begin{equation}\label{eq:operators}
	\begin{aligned}
		\M{O}_\mathrm{PS}&=\sqrt{\frac{1-T_{\mathrm{b}}}{T_{\mathrm{b}}}}\M{A}
		\sqrt{T_{\mathrm{b}}}^{\M{A}^\dagger\M{A}},\\
		\M{O}_\mathrm{PA}&=
		-\sqrt{1-T_{\mathrm{b}}}\M{A}^\dagger
		\sqrt{T_{\mathrm{b}}}^{\M{A}^\dagger\M{A}},\\
		\M{O}_\mathrm{PC}&=
		\sqrt{T_{\mathrm{b}}}\left(\frac{T_{\mathrm{b}}-1}{T_{\mathrm{b}}} \M{A}^\dagger\M{A} + 1 \right)
		\sqrt{T_{\mathrm{b}}}^{\M{A}^\dagger\M{A}}.
	\end{aligned}
\end{equation}
Suppose a non-Gaussian operation $\M{O} \in \lbrace \M{O}_\mathrm{PA},\M{O}_\mathrm{PS},\M{O}_\mathrm{PC} \rbrace$ is to be performed to a state.
Let $\M{\Xi}_{\mathrm{in}}$ be the density operator of the state.
The resultant state after the operation can be written as
\begin{equation}
	\M{\Xi}_{\mathrm{out}}=\frac{1}{\mathcal{N}}\M{O}\M{\Xi}_{\mathrm{in}}\M{O}^\dagger,
\end{equation}
where $\mathcal{N}=\operatorname{Tr}\left\{\M{O}\M{\Xi}_{\mathrm{in}}\M{O}^\dagger\right\}$ is a normalization constant, which is also the probability of success of the non-Gaussian operation.

%--------------------------------------
%
%Subtraction-then-addition:
%\begin{equation}
%\frac{T_{\mathrm{b}}-1}{T_{\mathrm{b}}} \M{A}^\dagger\M{A}
%T_{\mathrm{b}}^{\M{A}^\dagger\M{A}}.
%\end{equation}
%
%Addition-then-subtraction:
%\begin{equation}
%\left(T_{\mathrm{b}}-1\right) \M{A}\M{A}^\dagger
%T_{\mathrm{b}}^{\M{A}^\dagger\M{A}}.
%\end{equation}

\subsection{CV teleportation protocol with non-Gaussian operations}
\begin{figure}
	\centering
	\includegraphics[width=1\linewidth]{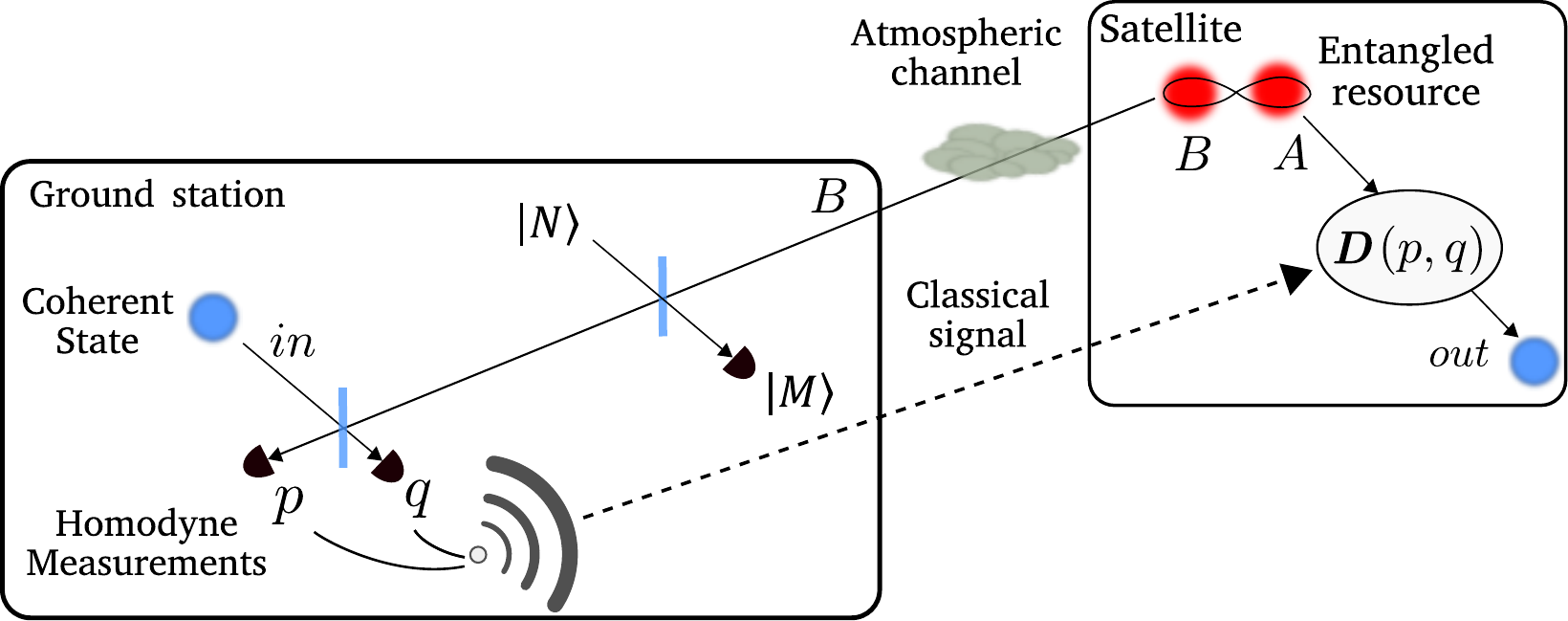}
	\caption{CV teleportation with non-Gaussian operations performed at the ground station.}
	\label{fig:diagngtp}
\end{figure}

In this section, we study the use of non-Gaussian operations in the protocol of CV quantum teleportation proposed by \cite{braustein1998teleportationCV}.
The deployment of the protocol over satellite channels has been discussed in previous sections, so we will only describe our modification in this section.
Our modified protocol is illustrated in Fig.~\ref{fig:diagngtp}, where we assume the satellite and the ground station already share some TMSV states that have been distributed over the noisy channel.
Before teleportation begins, the ground station will perform non-Gaussian operations to the local mode stored at the station.
The resultant non-Gaussian states shared between the satellite and the  ground station  will be used as the entangled resource for teleportation.

Similar to before, we will use the fidelity given by Eq.~(\ref{eq:fidelity}) as the metric to evaluate the effectiveness of our modified CV teleportation protocol.
%Recall, the fidelity can be written as
%\begin{equation}
%	\mathcal{F}=\frac{1}{\pi} \int \bigchi_{\mathrm{in}}(\xi) \bigchi_{\mathrm{out}}(-\xi) \, d^{2} \xi,
%\end{equation}
%where $d^{2} \xi=d\text{Re}({\xi})d\text{Im}({\xi})$, and $\bigchi_{\mathrm{in}}(\xi)$ and $\bigchi_{\mathrm{out}}(\xi)$ are the CFs of the input and output states of teleportation, respectively.
%The CFs satisfy the relation given by Eq.~(\ref{eq:marian}).
%In the absence of detection noise, the CFs satisfy
%\begin{equation}\label{eq:inoutnonoise}
%	\bigchi_{\mathrm{out}}(\xi)=\bigchi_{\mathrm{in}}(g\xi)\bigchi_{\mathrm{AB}}(\xi,\,g\xi^*),
%\end{equation}
%where again $g$ is the gain parameter for the teleportation protocol and $\bigchi_{\mathrm{AB}}(\xi_{\mathrm{A}},\xi_{\mathrm{B}})$ is the CF of the entangled resource state.
To determine the fidelity we need to derive the CFs of the non-Gaussian states.
We begin the derivation with the CF of the entangled state $\M{\Xi}$ shared between the ground station and the satellite.
This CF, which we repeat here for completeness, can be written as
%\begin{equation}
%	\begin{aligned}
%		\bigchi_{\mathrm{TMSV}}(\xi_{\mathrm{A}}, \xi_{\mathrm{B}})=\exp \bigg[-\frac{\left(1+\lambda^{2}\right)}{2\left(1-\lambda^{2}\right)}&\left(|\xi_{\mathrm{A}}|^{2}+|\xi_{\mathrm{B}}|^{2}\right).\\
%		+\frac{\lambda}{\left(1-\lambda^{2}\right)}&
%		\left(\xi_{\mathrm{A}} \xi_{\mathrm{B}}+\xi_{\mathrm{A}}^{*} \xi_{\mathrm{B}}^{*}\right)\bigg],
%	\end{aligned}
%\end{equation}
%where $\lambda=\tanh (r)$ with $r$ again being the squeezing parameter.
%Having passed through the channel, the CF of the entangled state shared between the satellite and the ground station, $\hat{\rho}$, is given by \cite{li2011time}
\begin{equation}
	\begin{aligned}
		\bigchi_{\mathrm{TMSV}}'(\xi_{\mathrm{A}}, \xi_{\mathrm{B}})=&\exp{\left[-\frac{1}{2}(\epsilon + 1 - T)|\xi_{\mathrm{B}}|^2\right]}\\
		&\times\bigchi_{\mathrm{TMSV}}(\xi_{\mathrm{A}},\sqrt{T}\xi_{\mathrm{B}}),
	\end{aligned}
\end{equation}
where again $\epsilon$ is the channel excess noise, $T$ is the channel transmissivity, and $\bigchi_\mathrm{TMSV}(\xi_{\mathrm{A}}, \xi_{\mathrm{B}})$ is the CF for the initial TMSV state prepared by the satellite -- which is given by Eq.~(\ref{eq:CF_TMSV}).
On performing PS to mode $B$ of $\M{\Xi}$, the \textit{unnormalized} CF of the resultant state is given by
\begin{equation}
	\begin{aligned}
		k_{\mathrm{PS}}(\xi_{\mathrm{A}}, \xi_{\mathrm{B}})
		=&\operatorname{Tr}\left\{\M{O}_{\mathrm{PS}}\M{\Xi}\M{O}_{\mathrm{PS}}^\dagger \M{D}(\xi_{\mathrm{A}})\M{D}(\xi_{\mathrm{B}})\right\}\\
		%&=p\operatorname{Tr}[\M{B}\sqrt{T_{\mathrm{b}}}^{\M{B}^\dagger\M{B}}\hat{\rho}\sqrt{T_{\mathrm{b}}}^{\M{B}^\dagger\M{B}}\M{B}^\dagger \hat{D}(\xi_{\mathrm{A}},\xi_{\mathrm{B}})]\\
		%=&p\operatorname{Tr}[\M{B}\hat{\rho}'\M{B}^\dagger \hat{D}(\xi_{\mathrm{A}},\xi_{\mathrm{B}})]\\
		=&\frac{T_{\mathrm{b}}-1}{T_{\mathrm{b}}}
		\exp{\left(\frac{|\xi_{\mathrm{B}}|^2}{2}\right)}\\
		&\times
		\frac{\partial^2}{\partial \xi_{\mathrm{B}} \partial \xi_{\mathrm{B}}^*}
		\bigg[
		\exp{\left(-\frac{|\xi_{\mathrm{B}}|^2}{2}\right)}
		f(\xi_{\mathrm{A}}, \xi_{\mathrm{B}}, \sqrt{T_{\mathrm{b}}})
		\bigg],
	\end{aligned}
\end{equation}
where
\begin{equation}
	\begin{aligned}
		f(\xi_{\mathrm{A}}, \xi_{\mathrm{B}}, \sqrt{T_{\mathrm{b}}})=&\int \frac{d \xi^2}{\pi (1-T_{\mathrm{b}})} \bigchi_{\mathrm{TMSV}}'(\xi_{\mathrm{A}}, \xi)\\
		&\times
		\exp{
			\bigg[
			\frac{1+T_{\mathrm{b}}}{2(T_{\mathrm{b}}-1)}(|\xi|^2+|\xi_{\mathrm{B}}|^2)}
		\bigg]
		\\
		&\times
		\exp{
			\bigg[
			\frac{\sqrt{T_{\mathrm{b}}}}{T_{\mathrm{b}}-1}
			(\xi_{\mathrm{B}}\xi^*+\xi_{\mathrm{B}}^*\xi)
			\bigg]},
	\end{aligned}
\end{equation}
and $\xi_{\mathrm{B}}$ and $\xi_{\mathrm{B}}^*$ are independent variables.

For PA and PC, the CF of the state after the non-Gaussian operations can be obtained in a similar fashion.
For PA the unnormalized CF is given by
\begin{equation}
	\begin{aligned}
		k_{\mathrm{PA}}(\xi_{\mathrm{A}}, \xi_{\mathrm{B}})
		=&(T_{\mathrm{b}}-1)
		\exp{\left(-\frac{|\xi_{\mathrm{B}}|^2}{2}\right)}\\
		&\times
		\frac{\partial^2}{\partial \xi_{\mathrm{B}} \partial \xi_{\mathrm{B}}^*}
		\bigg[
		\exp{\left(\frac{|\xi_{\mathrm{B}}|^2}{2}\right)}
		f(\xi_{\mathrm{A}}, \xi_{\mathrm{B}}, \sqrt{T_{\mathrm{b}}})
		\bigg].
	\end{aligned}
\end{equation}
For PC, the unnormalized CF is more involved, and is given by
\begin{equation}
	\begin{aligned}
		k_{\mathrm{PC}}(\xi_{\mathrm{A}}, \xi_{\mathrm{B}})
		=&q^2
		\exp{\left(\frac{|\xi_{\mathrm{B}}|^2}{2}\right)}
		\frac{\partial^2}{\partial \xi_{\mathrm{B}} \partial \xi_{\mathrm{B}}^*}
		\bigg\lbrace
		\exp{\left(-|\xi_{\mathrm{B}}|^2\right)}\\
		&\times
		\frac{\partial^2}{\partial \xi_{\mathrm{B}} \partial \xi_{\mathrm{B}}^*}
		\bigg[
		\exp{\left(\frac{|\xi_{\mathrm{B}}|^2}{2}\right)}
		f(\xi_{\mathrm{A}}, \xi_{\mathrm{B}}, \sqrt{T_{\mathrm{b}}})
		\bigg]\bigg\rbrace\\
		&-q
		\exp{\left(\frac{|\xi_{\mathrm{B}}|^2}{2}\right)}
		\frac{\partial}{\partial \xi_{\mathrm{B}}}
		\bigg\lbrace
		\exp{\left(-\xi_{\mathrm{B}}|^2\right)}\\
		&\times
		\frac{\partial}{\partial \xi_{\mathrm{B}}^*}
		\bigg[
		\exp{\left(\frac{|\xi_{\mathrm{B}}|^2}{2}\right)}
		f(\xi_{\mathrm{A}}, \xi_{\mathrm{B}}, \sqrt{T_{\mathrm{b}}})
		\bigg]\bigg\rbrace\\
		&-q
		\exp{\left(\frac{|\xi_{\mathrm{B}}|^2}{2}\right)}
		\frac{\partial}{\partial \xi_{\mathrm{B}}^*}
		\bigg\lbrace
		\exp{\left(-|\xi_{\mathrm{B}}|^2\right)}\\
		&\times
		\frac{\partial}{\partial \xi_{\mathrm{B}}}
		\bigg[
		\exp{\left(\frac{|\xi_{\mathrm{B}}|^2}{2}\right)}
		f(\xi_{\mathrm{A}}, \xi_{\mathrm{B}}, \sqrt{T_{\mathrm{b}}})
		\bigg]\bigg\rbrace\\
		&+f(\xi_{\mathrm{A}}, \xi_{\mathrm{B}}, \sqrt{T_{\mathrm{b}}}),
	\end{aligned}
\end{equation}
where $q=\frac{T_{\mathrm{b}}-1}{T_{\mathrm{b}}}$.

Additionally, we also investigate the sequential use of PS and PA.
We assume the two non-Gaussian operations adopt the same beam-splitter transmissivity.
For the scenario of PS followed by PA (PS-PA), the unnormalized CF is given by
\begin{equation}
	\begin{aligned}
		&k_{\mathrm{PS-PA}}(\xi_{\mathrm{A}}, \xi_{\mathrm{B}})\\
		&\quad=q^2
		\exp{\left(\frac{|\xi_{\mathrm{B}}|^2}{2}\right)}
		\frac{\partial^2}{\partial \xi_{\mathrm{B}} \partial \xi_{\mathrm{B}}^*}
		\bigg\lbrace
		\exp{\left(-|\xi_{\mathrm{B}}|^2\right)}\\
		&\quad\quad\times
		\frac{\partial^2}{\partial \xi_{\mathrm{B}} \partial \xi_{\mathrm{B}}^*}
		\bigg[
		\exp{\left(\frac{|\xi_{\mathrm{B}}|^2}{2}\right)}
		f(\xi_{\mathrm{A}}, \xi_{\mathrm{B}}, {T_{\mathrm{b}}})
		\bigg]\bigg\rbrace.\\
	\end{aligned}
\end{equation}
The unnormalized CF for PA followed by PS (PA-PS) is given by
\begin{equation}
	\begin{aligned}
		&k_{\mathrm{PA-PS}}(\xi_{\mathrm{A}}, \xi_{\mathrm{B}})\\
		&\quad=(T_{\mathrm{b}}-1)^2
		\exp{\left(-\frac{|\xi_{\mathrm{B}}|^2}{2}\right)}
		\frac{\partial^2}{\partial \xi_{\mathrm{B}} \partial \xi_{\mathrm{B}}^*}
		\bigg\lbrace
		\exp{\left(|\xi_{\mathrm{B}}|^2\right)}\\
		&\quad\quad\times
		\frac{\partial^2}{\partial \xi_{\mathrm{B}} \partial \xi_{\mathrm{B}}^*}
		\bigg[
		\exp{\left(-\frac{|\xi_{\mathrm{B}}|^2}{2}\right)}
		f(\xi_{\mathrm{A}}, \xi_{\mathrm{B}}, {T_{\mathrm{b}}})
		\bigg]\bigg\rbrace.\\
	\end{aligned}
\end{equation}

The \textit{normalized} CFs after the non-Gaussian operations are given by
\begin{equation}\label{eq:nGstate}
	\begin{aligned}
		\bigchi_{\text{x}}(\xi_{\mathrm{A}}, \xi_{\mathrm{B}})&=\frac{1}{k_{\text{x}}(0, 0)}k_{\text{x}}(\xi_{\mathrm{A}}, \xi_{\mathrm{B}}),\\
	\end{aligned}
\end{equation}
where $\text{x}\in{\left\lbrace \mathrm{PS}, \mathrm{PA}, \mathrm{PC}, \mathrm{PS-PA}, \mathrm{PA-PS}  \right\rbrace }$.
For compactness, the expressions for the CFs above are not shown here.

\subsection{Results}
\begin{figure}
	\centering
	\includegraphics[width=.95\linewidth]{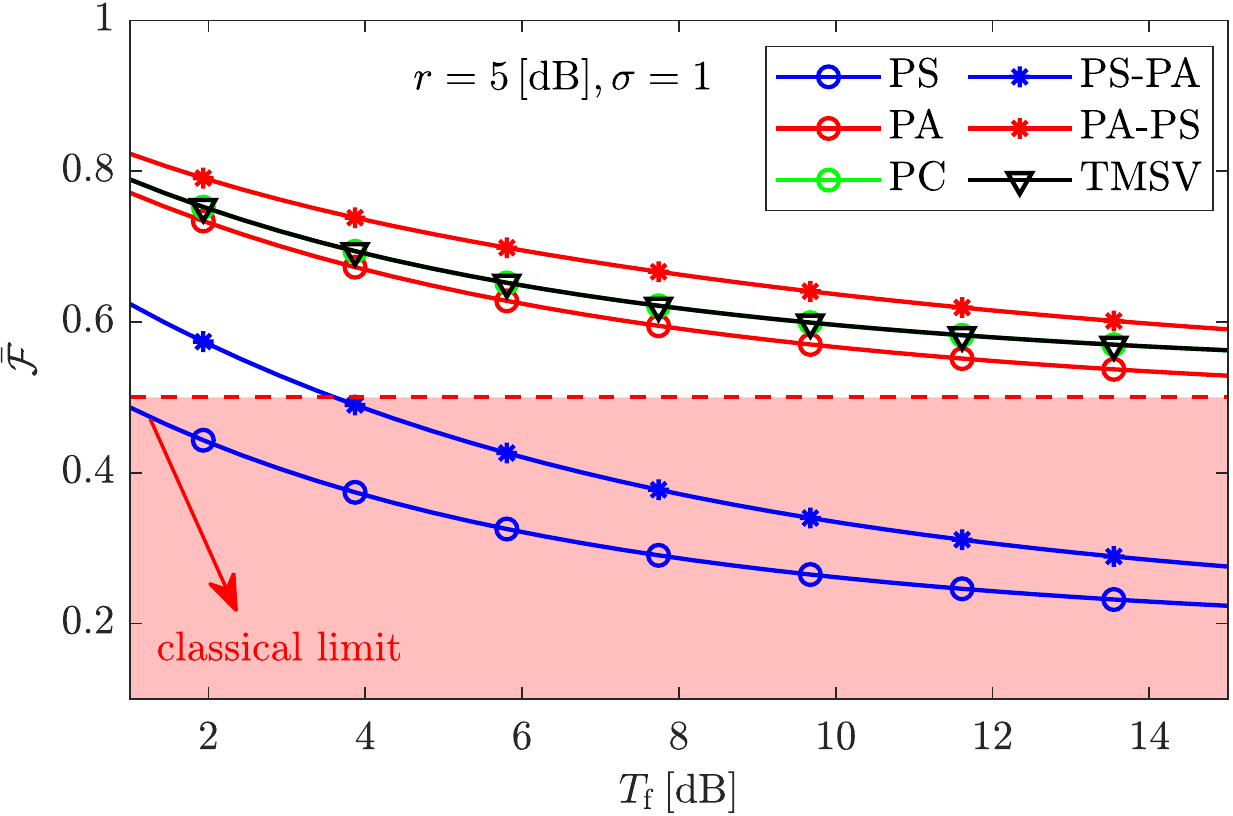}\\
	\includegraphics[width=.95\linewidth]{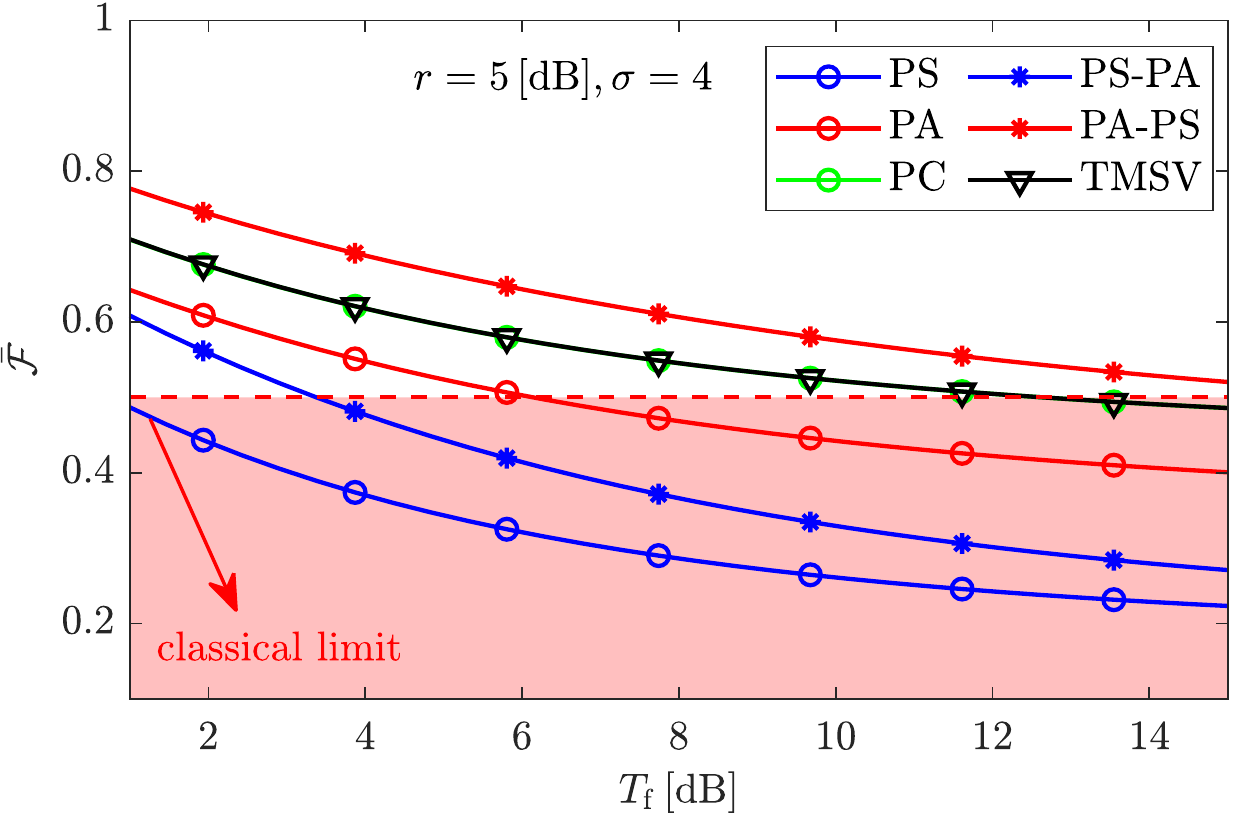}
	\includegraphics[width=.95\linewidth]{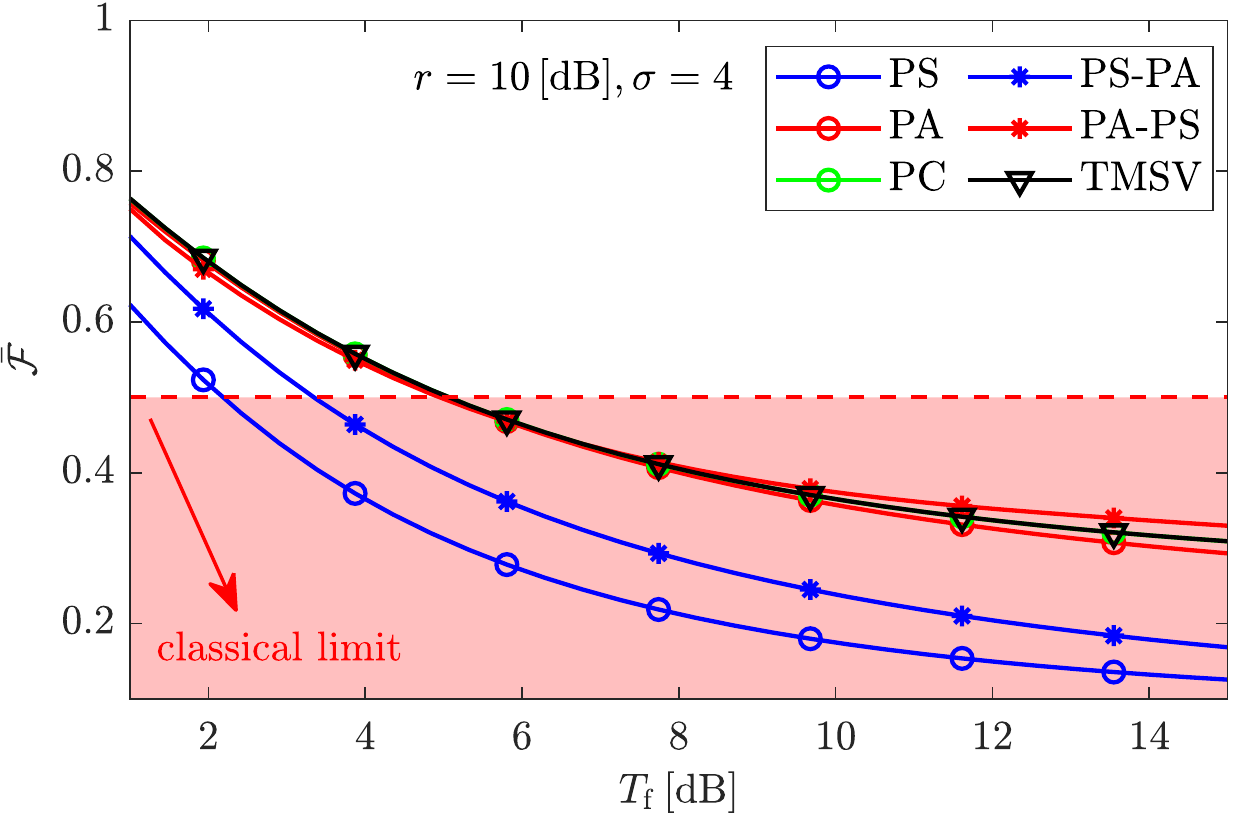}
	\caption{Mean fidelity \emph{vs.} effective channel transmissivity, where $r$ is the squeezing level (in dB) of the TMSV state prepared by the satellite, and $\sigma$ is the displacement variance of the input coherent states.
		The shaded area in red indicates the region where the teleportation fidelity is achievable by classical communications only.
		The effective excess noise is set according to Fig.~\ref{fig:channel} for $T_{\mathrm{f}}\geq7$ dB and is $1.4\cosh(r)\times10^{-2}$ otherwise. Recall, a higher ${T}_f$ in dB corresponds to higher loss.}
	\label{fig:figcoh}
\end{figure}

We study the teleportation of coherent states using non-Gaussian entangled resource states, of which the CF is chosen from Eq.~(\ref{eq:nGstate}) depending on which non-Gaussian operation is performed to the mode at the ground station.
We use the mean fidelity $\mathcal{\bar{F}}$ given by Eq.~(\ref{eq:avefidelity}) as our performance metric.
We adopt the effective channel loss and the effective excess noise obtained from the phase screen simulations (see Fig~\ref{fig:channel}).

In Fig.~\ref{fig:figcoh} we compare the maximized $\mathcal{\bar{F}}$ offered by various non-Gaussian operations against the effective channel loss $T_{\mathrm{f}}\,\mathrm{[dB]}$.
At each effective channel loss level, the maximization of $\mathcal{\bar{F}}$ is performed on the parameter space consisting of the transmissivity $T_{\mathrm{b}}$ of the beam-splitter in the non-Gaussian operations and the  gain parameter $g$ of the teleportation protocol.
For comparison, the case without any non-Gaussian operation is also included (i.e., the black curve in the figure).
In each sub-figure, $r$ is the squeezing parameter of the TMSV state generated by the satellite and $\sigma$ is the variance for the distribution of the displacement of the input coherent state (defined by Eq.~(\ref{eq:probcoherent})).
For $r$ the conversion from the linear domain to the dB domain is given by $r\,\mathrm{[dB]}\approx 8.67r$.
We see from Fig.~\ref{fig:figcoh} that among the five non-Gaussian operations we have considered, only PA-PS provides enhancement in $\mathcal{\bar{F}}$.
PA always provides larger $\mathcal{\bar{F}}$ than PS.
When $r$ is 5 dB, PA-PS provides the largest $\mathcal{\bar{F}}$ over the entire range of effective channel loss we have considered.

\begin{figure}
	\centering
	\includegraphics[width=.95\linewidth]{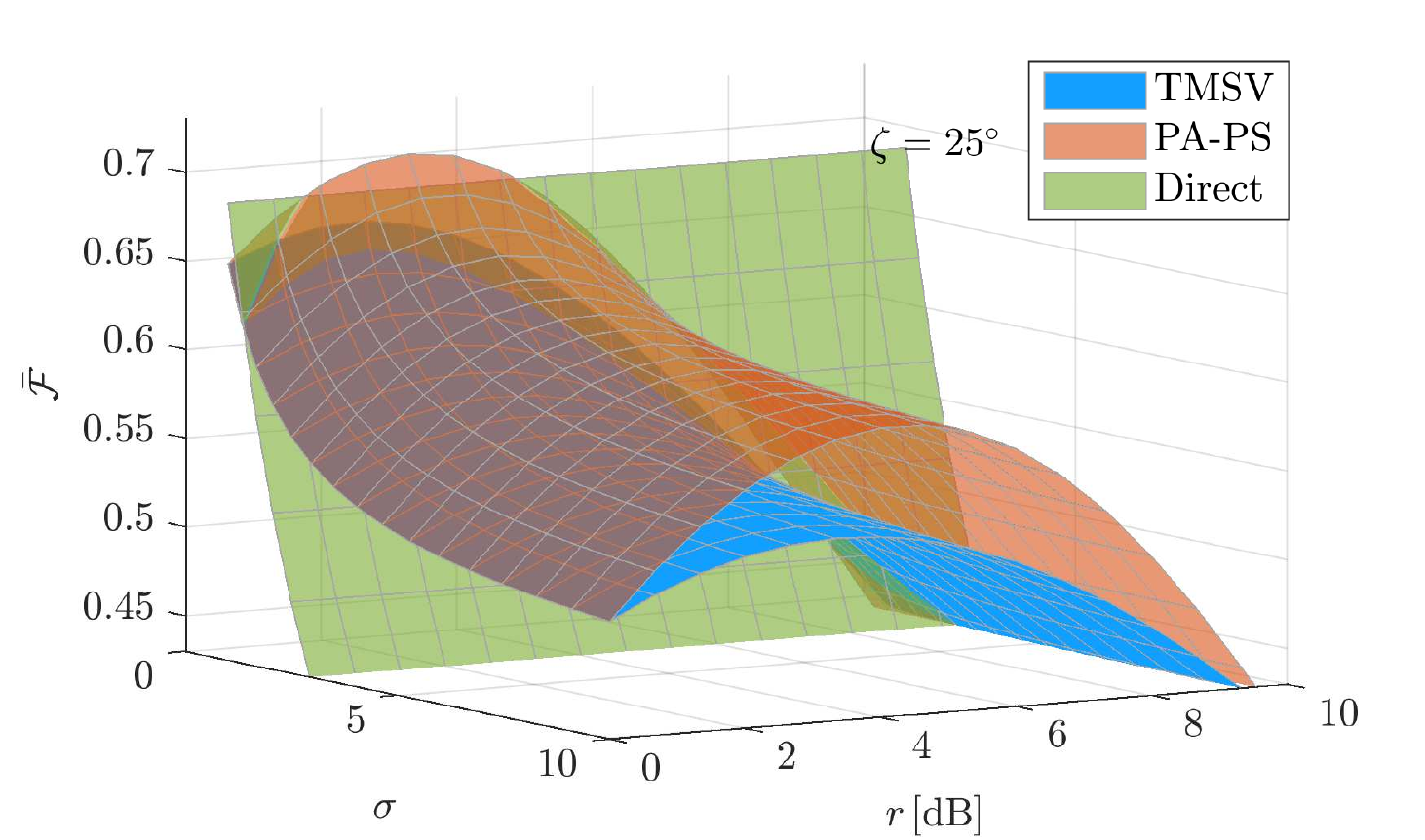}
	\includegraphics[width=.95\linewidth]{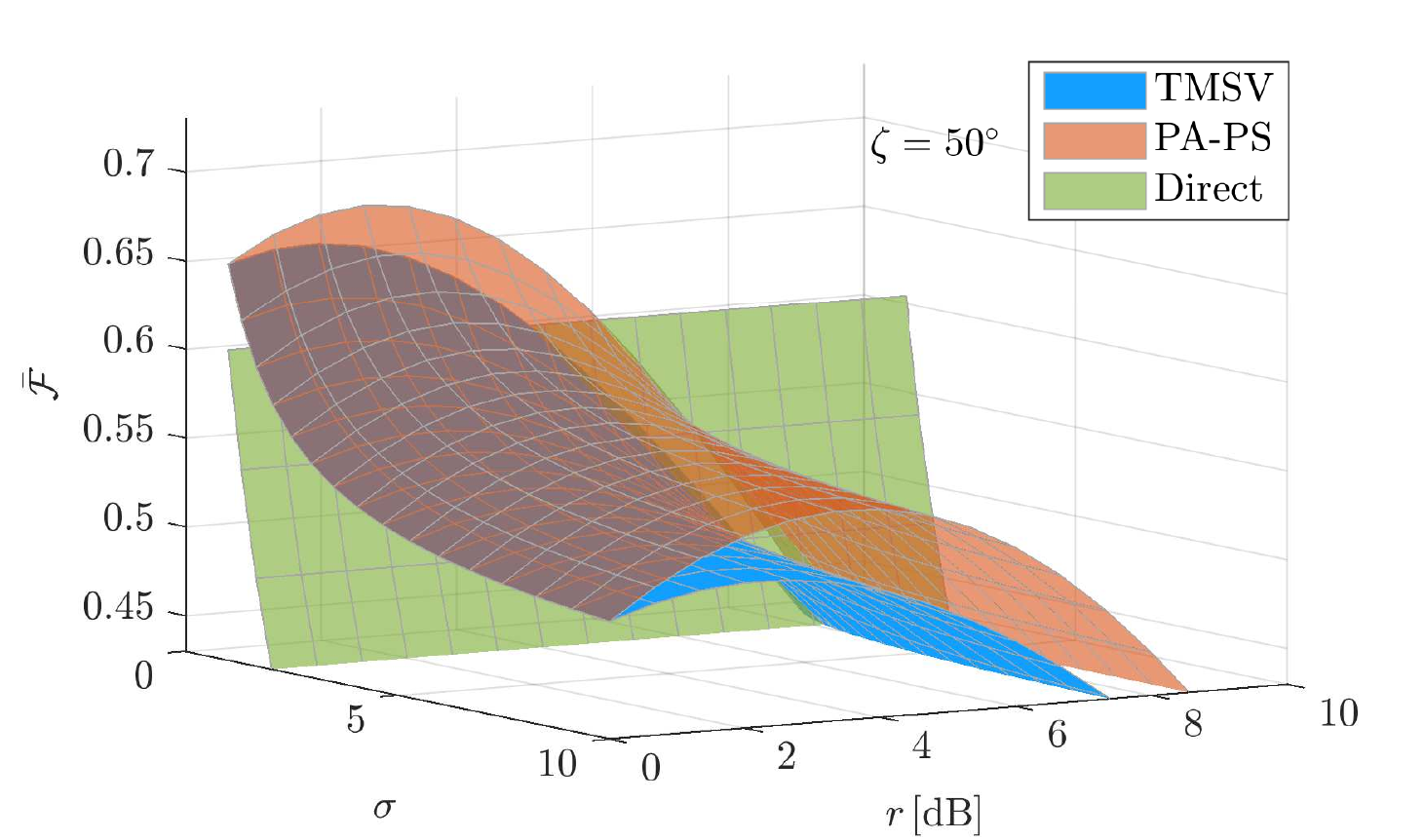}
	\caption{Mean fidelity as a function of the displacement variance of the coherent states $\sigma$ and the squeezing parameter $r$ of the TMSV state generated by the satellite. $\zeta$ is the satellite zenith angle.}
	\label{fig:figcohminr}
\end{figure}

We next compare the teleportation scheme with the non-Gaussian operation that provides the most improvement. That is, teleportation with PA-PS  compared to the direct transmission scheme.
The mean fidelity for the direct transmission scheme is given by Eqs.~(\ref{eq:dtcoherent}) and (\ref{eq:avefidelity}).
The results are illustrated in Fig.~\ref{fig:figcohminr},
where we compare the maximized $\mathcal{\bar{F}}$ against $r$ and $\sigma$ for different satellite zenith angles $\zeta$.
Again the maximization of $\mathcal{\bar{F}}$ is performed over the parameter space of $\left\lbrace T_{\mathrm{b}}, g \right\rbrace$.
We see in comparison to the original teleportation scheme (i.e. the TMSV case), the scheme with PA-PS can achieve the highest $\mathcal{\bar{F}}$ for the entire range of $\sigma$ we have considered.
PA-PS can also reduce the requirement on $r$ of the TMSV state prepared by the satellite (to reach a certain level of fidelity).
We also notice that when $\sigma$ is fixed, $\mathcal{\bar{F}}$ provided by the original teleportation scheme decreases when $r$ exceeds a certain value. The same trend is observed for the PA-PS scheme.

 In summary, we have shown in this section how non-Gaussian operations at the ground station can enhance the fidelity for teleporting coherent states by up to $10\%$. In addition,
 using such non-Gaussian operations, we have shown how the demand on the squeezing of the TMSV state prepared by the satellite can be reduced.

%%%%%%%%%%%%%%%%%% Discussion %%%%%%%%%%%%%%%%%%

\section{Discussion}
The focus of the present work is the use of CV teleportation channels for the teleportation of coherent states, and the use of non-Gaussian operations to enhance the communication outcomes. However, it is perhaps worth briefly discussing the flexibility of our system in regard to the transfer of other quantum states in the uplink, and the use of additional quantum operations. It will also be worth discussing differences and advantages of our system relative to DV-only systems - after all the only currently-deployed qiantum satellite system is one solely based on DV states \cite{Ren2017}.

\subsection{Other Quantum States and Operations}
Our scheme is actually applicable to any type of quantum state - even DV based systems. Some DV systems, e.g. polarization,\footnote{It is straightforward to alter polarization encoding into number-basis encoding or other forms of qubit encoding, e.g. \cite{Sychev2018}.} may need to be transformed first into the number basis.
 In number-basis qubit-encoding, vacuum contributions enter directly, similar to what we have discussed earlier. In such schemes, the use of the TMSV entangled teleportation channel (a CV channel) can be utilized as the resource to teleport the DV qubit state \cite{Lie2019}, and so our proposed scheme operates directly. Our scheme also operates directly on more complex quantum states such as hybrid DV-CV entangled states - even on both components of such states \cite{Hung}. This flexibility of CV entanglement channels over DV entanglement channels is another advantage offered by our scheme.

We also note, the non-Gaussian operations we have considered in this work represent a form of CV entanglement distillation \cite{Hage2008}. There are, of course, many other forms of CV entanglement distillation we could have considered at the ground receiver (or on-board the satellite) - we have only investigated  the simplest-to-deploy quantum operations.\footnote{Classical pre-processing via pre-selection based on transmissivity estimation using classical beams or post-processing based on measurement outcomes may assist these operations \cite{NedaReview}.} As technology matures (e.g. the advent of quantum memory), more sophisticated quantum operations (and entangled resources) will become viable as a means of further enhancing teleported uplink quantum communications; most likely out-competing any advances in the uplink-tracking technology that could assist direct communication. In principal, the  teleportation fidelity could approach one.

\subsection{DV Polarization - Micius}

We discuss now, known results from the LEO Micius satellite in the context of teleportation of DV-polarization states from the   ground to the satellite \cite{Ren2017}. Different from our system model, the teleportation experiment reported in \cite{Ren2017} does not use the downlink to create the entanglement, but rather utilizes the uplink as a means of distributing the entanglement. Therefore, the advantage of using the superior downlink channel is not afforded to that experiment. From the aperture used in \cite{Ren2017} - a 6.5cm radius transmitter and a 15cm radius receiver telescope, a turbulence induced loss of 30dB is obtained at a 500km altitude, the zenith distance of Micius. This translates into a beam width of 10m at the receiver plane (30m beam width and 40dB losses at 1400km is also reported). Nonetheless, the experiment still clearly demonstrates a fidelity of 0.8 for the teleportation of single-qubit encoded in single-polarized photons (well above the classical fidelity limit of 2/3 for a qubit), proving the viability of teleportation over the large distances tested.

In the context of the main idea presented in this work, use of the downlink channel (to create the entanglement channel) in an experimental set up similar to \cite{Ren2017} would be beneficial mostly in the context of an increased rate of teleportation, rather than an increase in fidelity. Our phase-screen simulations suggest (reversing the aperture sizes for a fair comparison, that is, 6.5cm radius transmitter at the satellite and 15cm radius on the receiving aperture) would result in a turbulence induced loss of 25dB, which would lead to a factor of $\sim 2-4$ enhancement in the teleportation rate relative to direct transmission. Of course, if we increase the ground receiver aperture, larger enhancements could be found. The fact that it is much easier to deploy large telescopes on the ground, compared to in space, is another advantage of our teleportation scheme.
%The reason for this is that in the teleportation of a polarization-encoded qubit, loss manifests itself largely as a lower detection rate - the vacuum contribution does not enter the teleportation channel as it does in a CV entangled channel.

Let us briefly outline the main differences in DV-polarization teleportation relative to CV teleportation.
 In DV-polarization implementations the vacuum contribution does not enter the teleportation channel in the same manner it does in a CV entangled channel. In the DV-polarization channel the loss enters our calculations primarily via two avenues. One avenue is simply through the different raw detection rates set by the differential evolution of the beam profiles in the downlink and uplink. As discussed, in the downlink the beam width at the receiver will be smaller than in the uplink. For a given receiver aperture this translates into an increased detection rate in and by itself. We can use the phase screen calculations described earlier (e.g. Fig.~\ref{fig:fig_T_CMP} for equal transmit and receive apertures of 1m) to determine this  rate increase.  The second avenue is a manifestation of the vacuum through dark counts in the photodetectors. In real-world deployments of teleportation through long free-space channels \cite{Ren2017,Yin2012,Ma2012} a coincidence counter is used to pair up entangled photons, typically with a time-bin width of 3ns \cite{Ren2017}. Due to the presence of a vacuum in almost all time-bins, only of order 1 in a million events are triggered as a photon-entangled pair. Dark counts in the best photodetectors are currently in the range of 20Hz. However, in orbit, and because of stray light, combined background counts are more likely to be of order 150Hz \cite{Ren2017}. A background count in one time-bin will lead to a false identification of an entangled pair generated between the satellite and ground station. This is different to the CV scenario where each time bin is assumed to contain a pulse - albeit one contaminated with a vacuum contribution.

Another major difference in DV \emph{vs.} CV teleportation systems is contamination caused by higher order terms in the production of the (single) photons that are to be teleported in the DV systems. The optimal probability of single-photon emission (set by the user) decreases with increasing loss \cite{Bourgin2013}. This is due to a lower probability leading to a reduction in the number of double pair emissions that lead to flawed Bell measurements.  This effect is counteracted by the strength of the source that emits the two-photon entangled pairs (set by the user) - the optimal value of which increases with increasing loss. These two parameters can be jointly optimised for the loss anticipated, leading to asymmetric parameter settings for the downlink and uplink teleportation deployment \cite{Bourgin2013}.   An additional issue relevant to DV-polarization teleportation is partial photon distinguishably at the Bell state measurement which leads to a drop in interference at the beam splitter, and, of course, polarization errors (in production or measurement).

%For a spontaneous parametric
%down-conversion source this is described by the relation
%${\left| \varphi  \right\rangle _{In,Tr}} = \left| 0 \right\rangle  + \sqrt \chi  (A_{In}^\dag A_{Tr}^\dag )\left| 0 \right\rangle  + \frac{\chi }{2}{(A_{In}^\dag A_{Tr}^\dag )^2}\left| 0 \right\rangle $, where the subscripts \emph{In} and \emph{Tr }refer to the input state that is to teleported and  the trigger photon, respectively. Here, $\chi$ is the single-photon emission probability, and as we can see the probability of double emission is $\chi^2$ \cite{Ren2017}.

The relative importance of all the above terms for free space teleportation from ground to satellite are considered to be background counts (4$\%$), higher-order photon emission (6$\%$), polarization errors (3$\%$), and  photon  indistinguishability (10$\%$) \cite{Ren2017}. In a series of experiments over 100km \cite{Ma2012}, 143km \cite{Ma2012} and ground-to-satellite \cite{Ren2017} a fidelity of teleportation in the range $0.8-0.9$ was obtained by all.

Another issue in discussing DV relative to CV teleportation is the classical teleportation fidelity of both systems. That is, the fidelity that can be achieved by purely classical information being communicated across the channel (e.g. the classical information representing the outcome of a particular quantum measurement). This classical information allows the receiver to partially reconstruct the desired quantum state. In the coherent state teleportation discussed earlier this classical fidelity was 1/2. However, for DV qubits it is 2/3. This fact translates into a less useful range of teleportation fidelity for the DV scenario relative to the CV scenario. Finally, it is worth noting that the Bell state measurements used currently in DV systems are only 50$\%$ efficient. This is a consequence of the fact that Bell state measurements based on linear optics can only discriminate between two of the four Bell states. Although, in principal, full Bell state measurements in the DV basis are possible (eg via ancilla and two-qubit interactions), no real-world implementation of the latter exist - all current deployments utilize a linear-optics-only solution \cite{Ren2017,Yin2012,Ma2012}.

\subsection{Future work}
We recognise other input states may lead to an enhanced fidelity in both the direct uplink transmission channel and via the resource CV teleportation channel. It is likely that in these circumstances we will again find some channel parameter settings where teleportation leads to better communication outcomes. However, coherent states and TMSV states are easy to produce and are considered the ``workhorses'' of CV quantum communications, and are therefore the focus of this work. We also recognize more sophisticated set-ups could be considered, such as the use of classical feedback on channel conditions to optimise the parameters of the input states (e.g. squeezing levels and amplitudes). However, such improvements are at the cost of a considerable increase in implementation complexity. Again, it is likely that in these circumstances some channel parameter settings will provide for communication gains via teleportation relative to direct transfer.  Future investigations that properly identify such channel settings would be useful. Our study has also been limited in terms of the aperture settings we have adopted. We have used what we consider to be aperture settings likely deployable in next-generation systems which take space-based quantum communication to the production phase. Further study of possible teleportation gains for a wider range of aperture settings would also be useful.

%%%%%%%%%%%%%%%%%% Conlcusions %%%%%%%%%%%%%%%%%%

 \section{Conclusions}
In this work, we have investigated the use of a CV teleportation channel, created between a LEO satellite and a terrestrial ground station, as a means to enhance quantum communication in uplink satellite communications. Such communications are expected to be very difficult in practice due to the severe turbulence-induced losses anticipated for uplink satellite channels. Our CV teleportation channel was modelled using the superior (lower loss) downlink channel from the satellite as a means to distribute one mode of an in situ satellite TMSV state to the terrestrial station - a form of long-range  entanglement distribution that may become mainstream in coming years. Our results showed that use of this teleportation channel for uplink coherent state transfer is likely to be much superior to coherent state transfer directly through the uplink channel. The use of non-Gaussian operations at the ground station was shown to further enhance this superiorly. Given the flexibility of CV teleportation as a means to invoke all forms of quantum state transfer beyond just coherent state transfer, it could well be the scheme introduced here could become the de facto choice for all future uplink quantum communication with satellites.

%%%%%%%%%%%%%%%%%% BIB %%%%%%%%%%%%%%%%%%

%\bibliography{mybib,References_Ziqing,bib}

\bibliography{bibcombined}

%\EOD

\end{document}